# Tracking the Ultraviolet Photochemistry of Thiophenone During and Beyond the Initial Ultrafast Ring Opening

Shashank Pathak[1*], Lea M. Ibele[2*], Rebecca Boll[3], Carlo Callegari[4], Alexander Demidovich[4], Benjamin Erk[5], Raimund Feifel[6], Ruaridh Forbes[7], Michele Di Fraia[4], Luca Giannessi[4,8], Christopher S. Hansen[9], David M.P. Holland[10], Rebecca A. Ingle[11], Robert Mason[12], Oksana Plekan[4], Kevin C. Prince[4,13], Arnaud Rouzée[14], Richard J. Squibb[6], Jan Tross[1], Michael N.R. Ashfold[15], Basile F.E. Curchod[2], and Daniel Rolles[1]

*both authors have contributed equally

[1]*J.R. Macdonald Laboratory, Department of Physics, Kansas State University, Manhattan, KS, USA*
[2]*Department of Chemistry, Durham University, Durham DH1 3LE, UK*
[3]*European XFEL, Schenefeld, Germany*
[4]*Elettra - Sincrotrone Trieste S.C.p.a., 34149 Basovizza, Trieste, Italy*
[5]*Deutsches Elektronen-Synchrotron, 22607 Hamburg, Germany*
[6]*Department of Physics, University of Gothenburg, Gothenburg, Sweden*
[7]*Stanford PULSE Institute, SLAC National Accelerator Laboratory, Menlo Park, CA 94025, USA*
[8]*Istituto Nazionale di Fisica Nucleare, Laboratori Nazionali di Frascati, 00044 Frascati (Rome), Italy*
[9]*School of Chemistry, University of New South Wales, Sydney NSW 2052, Australia*
[10]*Daresbury Laboratory, Daresbury, Warrington, Cheshire WA4 4AD, UK*
[11] *Department of Chemistry, University College London, London, WC1H 0AJ, UK*
[12]*Department of Chemistry, University of Oxford, Oxford, OX1 3TA, UK*
[13]*Centre for Translational Atomaterials, Swinburne University of Technology, Melbourne, Australia*
[14]*Max-Born-Institut, 12489 Berlin, Germany*
[15]*School of Chemistry, University of Bristol, Bristol, BS8 1TS, UK*

**Abstract**

Photo-induced isomerization reactions, including ring-opening reactions, lie at the heart of many processes in nature. The mechanisms of such reactions are determined by a delicate interplay of coupled electronic and nuclear dynamics unfolding on the femtosecond scale, followed by the slower redistribution of energy into different vibrational degrees of freedom. Here we apply time-resolved photoelectron spectroscopy with a seeded extreme ultraviolet free-electron laser to trace the ultrafast ring opening of gas-phase thiophenone molecules following photoexcitation at 265 nm. When combined with cutting-edge *ab initio* electronic structure and molecular dynamics calculations of both the excited- and ground-state molecules, the results provide unprecedented insights into both electronic and nuclear dynamics of this fundamental class of reactions. The initial ring opening and non-adiabatic coupling to the electronic ground state is shown to be driven by ballistic S–C bond extension and to be complete within 350 femtoseconds. Theory and experiment also allow clear visualization of the rich ground-state dynamics – involving formation of, and interconversion between, several ring-opened isomers and the reformed cyclic structure, and fragmentation (CO loss) over much longer timescales.

**Introduction**

Recent advances in time-resolved experimental techniques and in computational methods for treating (coupled) electronic and nuclear dynamics are revolutionizing the field of ultrafast photochemistry, enabling direct probing of evolving molecular structures with unprecedented structural and temporal resolution[1-8]. Such studies provide the ultimate test of our knowledge and understanding of light-initiated chemistry. Photo-induced ring-opening/closing reactions play a crucial role in many key processes in nature, *e.g.*, in the synthesis of natural products (such as previtamin $D_3$ by sunlight), and are currently attracting interest as molecular and biomolecular switches for photo-controlled switching of enzyme activity, optical data storage, modulating energy and electron transfer processes[9,10] and in potential medical applications[11]. Ring-opening reactions featured prominently in the development of the Woodward-Hoffmann rules that help rationalize the mechanisms and outcomes of pericyclic reactions. It is now recognized that photo-induced isomerization (including ring-opening) reactions are governed by strong non-adiabatic coupling between multiple electronic states within the molecule via conical intersections, which represent a breakdown of the Born-Oppenheimer approximation[5,12].

The photo-induced ring opening of the polyene 1,3-cyclohexadiene[13] is widely employed as a model system for benchmarking and validating ultrafast methods, *e.g.* by ultrafast X-ray[1,7,8] and electron diffraction[6], by femtosecond transient X-ray absorption[4] and fragmentation[14], and by time-resolved photoelectron spectroscopy (TRPES)[15-16]. However, few other photo-induced ring-opening reactions have been probed so thoroughly and, of these, even fewer have provided a comprehensive picture of the reaction dynamics on both the excited and ground ($S_0$) state potential energy surfaces (PESs).

Here we report a combined theoretical and experimental study of the ultraviolet (UV) photo-induced ring opening of a prototypical heterocyclic molecule, 2(5H)-thiophenone ($C_4H_4OS$, henceforth thiophenone; see Fig. 1). Heterocyclic compounds are fundamental building blocks in the synthesis of many organic compounds. Studying these "single units" may help in validating the (necessarily more complex and less resolvable) photochemistry of ever larger molecules. The study is conducted in the gas phase (*i.e.* under collision-free conditions) and thus reveals information on the purely intramolecular relaxation pathways, without the solvation effects present in previous matrix isolation[17] and liquid-phase[18] studies of this system. Theory and experiment combine to afford detailed insights into both the mechanism and time scale of the initial ring-opening process, and the subsequent evolution of the vibrationally excited ground-state photoproducts.

The experimental study employs the extreme ultraviolet (XUV) radiation provided by the free-electron laser (FEL) FERMI[19]. TRPES[20,21] is sensitive to both electronic and structural dynamics – which is essential for any full understanding of the coupled electronic and nuclear dynamics that govern most photo-induced reactions. TRPES also allows access to so-called 'dark' states that may not be amenable to study by transient absorption methods. XUV probe pulses (from an FEL or a high harmonic generation source) are sufficiently energetic to ionize molecules in the electronic ground state ($S_0$) – thereby overcoming a long-recognized shortcoming of TRPES studies that employed lower energy (UV) photons and were thus unable to reveal the ultimate structural dynamics following transfer from an electronically excited state to the ground state[16,22-26]. FERMI is a seeded FEL[19,27], in which an external laser is used to initiate the XUV generation, offering the advantage of (i) a narrower photon energy bandwidth, *i.e.* higher energy resolution, and higher stability compared to FELs based on self-amplified spontaneous emission (SASE), and (ii) higher pulse energies and photon fluxes as compared to monochromatized XUV sources based on high harmonic generation.

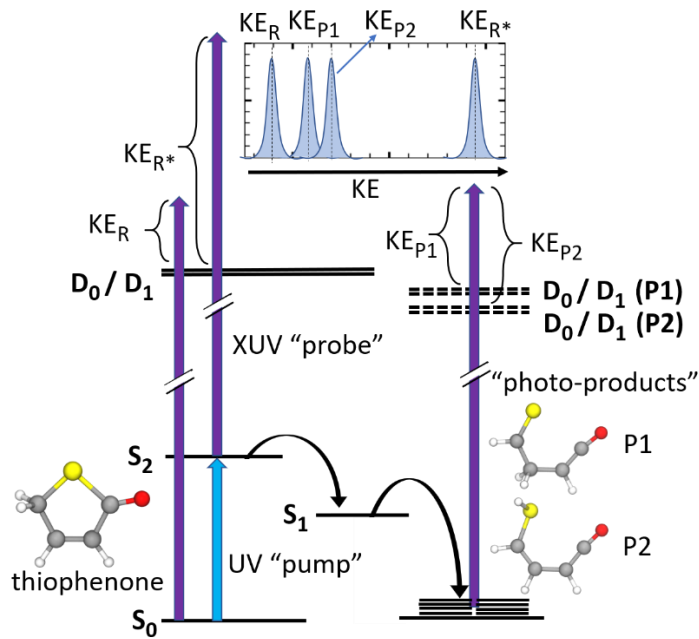

**Figure 1: Schematic of the UV excitation, ring opening and photoionization of thiophenone.** The molecule is photoexcited from its ring-closed ground state ($S_0$) to an electronically excited state ($S_2$). It evolves through an optically dark excited state ($S_1$) back to the (vibrationally excited) ground electronic state of several possible reaction products (P1, P2, *etc*). The XUV probe photon energy is sufficient to ionize thiophenone and all reaction products from both ground and excited states into several ionic final states ($D_0$, $D_1$, etc.). The time-evolving electron kinetic energy (KE) spectrum (top) thus consists of contributions from the ground and excited states of thiophenone (denoted as R and R*, respectively) and from the different products. In the depictions of the molecular geometry, carbon atoms are shown in grey, hydrogen in white, oxygen in red, and sulfur in yellow.

In the present experiment, gas-phase thiophenone molecules are excited to the optically bright $S_2$ electronic state by a UV pulse ($\lambda \approx 265$ nm). The excited molecules evolve through one or more conical intersections at progressively greater C–S bond separations *en route* back to the $S_0$-state PES, where they can adopt the original closed-ring or several possible open-ring geometries[18,28], some of which are sketched in Fig. 1. The evolving wavepacket and the formation of open- and closed-ring photoproducts are probed by ionizing the molecule with a time-delayed 19.24-eV ($\lambda$ = 64.44 nm) XUV pulse. Time-dependent photoelectron time-of-flight spectra are recorded as a function of time delay ($\Delta t$) between the UV and FEL pulses using a magnetic-bottle spectrometer (see Methods and Supplementary Material (SM) Fig. S1). The FEL photon energy is deliberately chosen to be well above the first ionization potential (IP) of the ground state molecule (~9.7 eV)[29] but below the IP of the helium carrier gas, ensuring that the electronic character of the target molecule can be traced throughout the complete structural evolution with only minimum background signal from the carrier gas. The experimental results are complemented by high level *ab initio* electronic structure and molecular dynamics calculations of both excited- and ground-state molecules, which afford unprecedented insight into the ultrafast electronic de-excitation that accompanies ring opening and the subsequent interconversion between different isomeric forms of the highly vibrationally excited ground-state photoproducts.

**Results and Discussion**

*Experimental Results.* The experiment measures electron time-of-flight spectra of thiophenone as a function of $\Delta t$, which are converted to electron kinetic energy spectra and then, by energy conservation, into spectra of the valence binding energies (see Methods). Fig. 2(a) shows such spectra in the form of a two-dimensional (2-D) plot of electron yield as functions of binding energy (BE) and $\Delta t$. The dominant feature at BE ~9.7 eV is due to photoionization of the "cold" (*i.e.* non-excited) closed-ring $S_0$-state thiophenone molecule[29]. This peak is depleted by ~20% for positive $\Delta t$, as shown in Fig. 2(b) (red circles), confirming excitation of ground-state molecules by the UV pulse. A fit to the delay-dependent yield of photoelectrons originating from $S_0$-state parent molecules yields an upper limit for the temporal instrument response function of $\sigma = 72 \pm 8$ fs. Fig. 2(a) reveals photoelectrons with BEs as low as ~5 eV at the shortest positive $\Delta t$. The prompt appearance and subsequent decay of this contribution is also emphasized in Fig. 2(b), which shows a Gaussian-shaped transient signal with $\sigma=76 \pm 6$ fs (blue triangles). With increasing delay, the signal at BE ~5 eV fades and the peak intensity shifts toward higher BE. As Fig. 2(c) shows, the peak in the intensity *vs* $\Delta t$ transient obtained by taking contiguous 0.6 eV-wide slices for BE $\geq$ 5.3 eV shifts to progressively later $\Delta t$ with increasing BE, and the transients gain an increasingly obvious tail.

The thiophenone cation has close-lying ground ($D_0$) and first excited ($D_1$) states at a vertical IP ($IP_{vert}$) of ~9.7 eV, and higher excited states at $IP_{vert}$ values of 10.58 eV ($D_2$), 12.25 eV ($D_3$) and 14.1 eV ($D_4$)[29]. Given the present pump photon energy of 4.67 eV, the signal appearing at a binding energy of ~5 eV at $\Delta t$ ~0 is readily attributed to vertical ionization of photoexcited molecules in the $S_2$ state to the (unresolved) $D_0$ and $D_1$ states. Ionization to the latter is strongly disfavored due to selection rules, as discussed in SM Section S2.3. The evolution of the signal at BE ~5 eV, the peak shift towards higher BE at later $\Delta t$, and the more intense tail in the $\Delta t$ traces for higher BE slices all reflect the complex evolution of the photo-prepared wavepacket, which the accompanying theory shows involves ultrafast depopulation of the $S_2$ state to yield "hot", *i.e.*, highly vibrationally excited, $S_0$ (henceforth $S_0^\#$) molecules (revealed by the green stripe in Fig. 2(a) at BE ~9 eV). Note that any photoelectrons arising from ionization of $S_0^\#$ molecules to excited $D_n$ ($n > 1$) cation states are likely to appear at BE >10 eV (see extended spectrum in SM Fig. S2), and thus do not affect the discussion that follows.

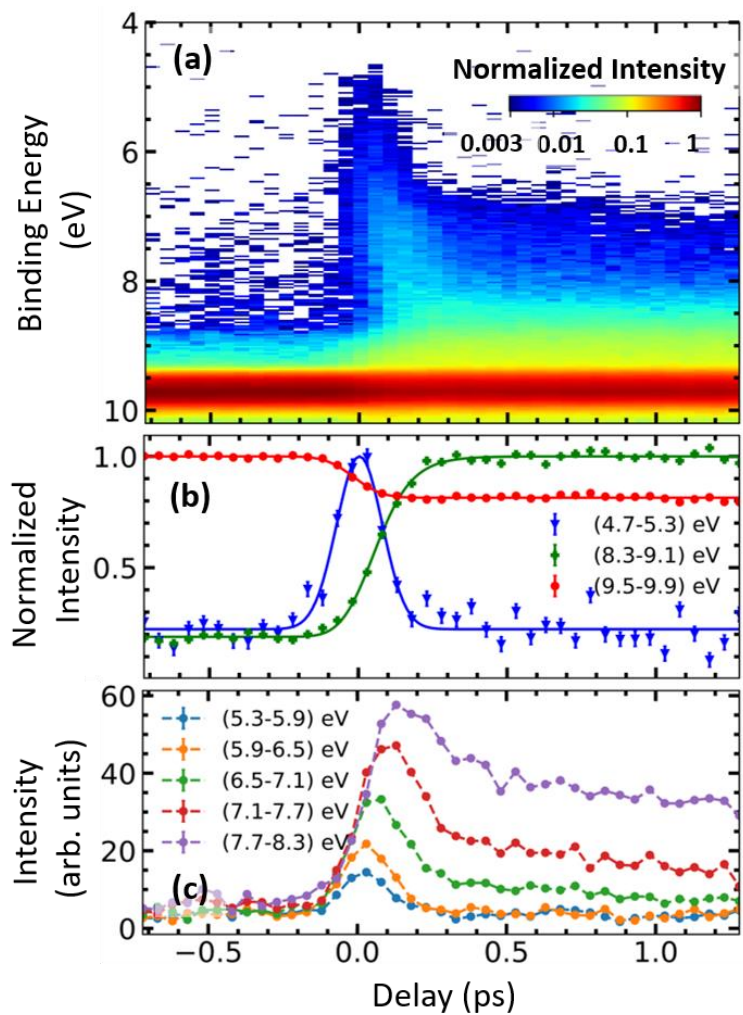

**Figure 2: Time-dependent photoelectron spectra of UV-excited thiophenone.** (a) Measured photoelectron yield as a function of binding energy (BE) and time delay ($\Delta t$) between pump and probe pulses (see SM Fig. S2 for a plot over a wider range of signal intensities, binding energies and delays). Negative $\Delta t$ corresponds to the FEL pulse preceding the UV pulse, positive $\Delta t$ to the UV pulse preceding the FEL pulse, while the color represents the photoelectron yield. Regions with an intensity below 0.003 are shown in white. Delay-dependent photoelectron yields for (b) three BE ranges selected to illustrate the photoinduced de-population of the $S_0$ state (red circles), population of the $S_2$ state (blue triangles), and the build-up of vibrationally excited $S_0^{\#}$ photoproducts (green crosses); and (c) five contiguous 0.6 eV-wide BE slices that inform on the evolution from photoexcited $S_2$ to internally excited $S_0^{\#}$ molecules. Statistical error bars are included but are generally smaller than the symbol size. The parameters of the least-square fits shown as solid lines in (b) are summarized in SM Table S1, while the dashed lines in (c) just join the dots for better visibility. The data in (b) have been normalized such that the maximum value of the fit is at 1 for each curve, whereas the data in (c) are displayed on a common intensity scale.

*Calculations of decay pathways and excited-state dynamics.* To interpret the dynamics revealed in the TRPES spectra, the lowest-lying PESs of thiophenone were computed and different critical points were located using SA(4)-CASSCF(10/8) calculations, and their energies were further refined using XMS(4)-CASPT2(10/8) (see Computational Details for more information). The Franck-Condon (FC) geometry corresponds to the equilibrium structure of the $S_0$-state of thiophenone, wherein the highest occupied molecular orbital (HOMO) is an out-of-plane $\pi$ orbital largely localised on the S atom, henceforth labelled $n(S)$. At our chosen pump photon energy, thiophenone is predominantly excited to its $S_2$ state. At the XMS(4)-CASPT2(10/8) level of theory, the $S_0 \rightarrow S_2$ transition has $n(S)/\pi^*$ character, a calculated transition energy of 4.67 eV and an appreciable oscillator strength (0.036 a.u.) – reflecting the constructive overlap of the donating $n(S)$ and accepting $\pi^*$ orbitals. The $S_0 \rightarrow S_1$ transition (with a calculated energy of 4.20 eV), in contrast, has $n(O)/\pi^*$ character and is dark (the donating $n(O)$ orbital lies in the plane of the ring and is thus orthogonal to the accepting orbital, see SM Section S2.1). Different minimum energy conical intersections (MECIs) were located between the $S_2$, $S_1$, and $S_0$ states of thiophenone, as shown in Fig. 3(a). All of these MECIs indicate a ring opening of thiophenone in the excited electronic state (*i.e.* formation of a biradical), followed by geometrical relaxation – *e.g.* twisting of the $CH_2$–S moiety out of the molecular plane and bending of the C–C=O moiety (see also SM Section S2.2).

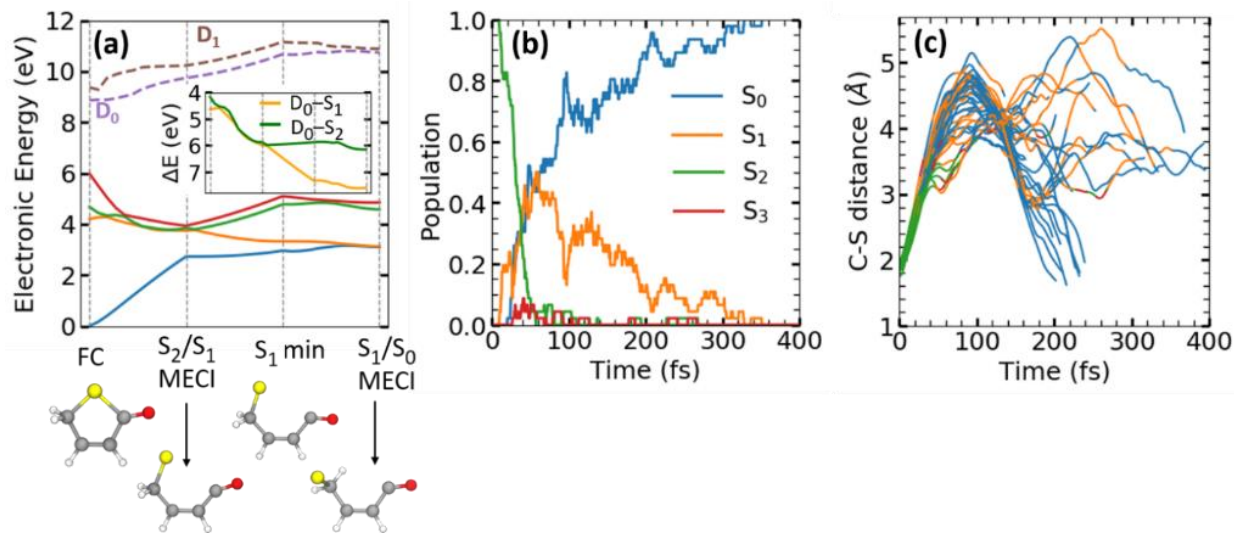

**Figure 3:** (a) PESs for the lowest neutral singlet (solid lines) and cationic doublet (dashed lines) electronic states of thiophenone along linear interpolation in internal coordinates (LIIC) pathways between different critical geometries. Electronic energies were obtained using XMS(4)-CASPT2(10/8) for the neutral and XMS(4)/CASPT2(9/8) for the cationic form of thiophenone. The critical geometries, minima, and minimum energy conical intersections (MECIs), located using the SA(4)-CASSCF(10/8) level of theory, are indicated with light grey vertical lines, and the geometries of these critical points on the PESs of neutral thiophenone are shown below. LIIC pathways have then been computed between each critical point. The inset shows the electronic energy gap $\Delta E$ between the $D_0$ state of the cation and the second ($S_2$) or first ($S_1$) singlet states of neutral thiophenone for each point along the LIIC pathway. (b) Time-dependence of the $S_3$, $S_2$, $S_1$ and $S_0$ state populations provided by the TSH dynamics (46 trajectories). (c) Time-evolution of the C−S bond distance for each of these 46 trajectories, illustrating the prompt initial bond extension in all cases and the trajectory-dependent evolution from $S_2$ through $S_1$ to the $S_0$ state (coded in the same color as the PESs in (a) and (b)). The TSH trajectories were propagated until they experienced electronic structure instabilities in the ground state (see text).

To determine the possible connections between these critical structures and the relaxation pathways of thiophenone following UV photoexcitation, linear interpolations in internal coordinates (LIICs) were performed using XMS(4)-CASPT2(10/8) (see Computational Details). Starting from the FC geometry, the LIICs smoothly connect the different critical points and confirm that photoexcited thiophenone ($S_2$) molecules can relax efficiently towards the $S_0$ state via the $S_2/S_1$ and $S_1/S_0$ seams of intersection in the C–S bond extension coordinate[28]. The energies of the low-lying (*i.e.* $D_0$ and $D_1$) states of the thiophenone cation have also been computed at selected geometries along the LIIC pathways (Fig. 3(a), dashed lines). In the Franck-Condon region, these states are characterized by removal of an electron from, respectively, the *n*(S) and *n*(O) lone pair orbitals. As Fig. 3(a) shows, the topographies of the various PESs of neutral thiophenone vary strongly along the LIIC pathways, but the energies of the $D_0$ and $D_1$ states of the cation show a smooth and very gradual increase. The energy differences ($\Delta E$) between the $D_1/D_0$ and the $S_2/S_1$ potentials along the LIIC pathways (inset in Fig. 3(a)) increase dramatically from the FC point out to the $S_1/S_0$ MECI. The calculations allow us to assign the experimentally observed rapid increase in BE with increasing $\Delta t$ to the ultrafast depopulation of the $S_2$ state and electronic deactivation to the $S_0$ state, resulting in highly vibrationally excited ground state molecules. We note that the calculated $\Delta E$ values are consistently slightly lower than the experimental BE values. Such underestimation of (experimental) IP values by (X)MS-CASPT2 methods is well-known[30] and, in the present case, can also be related to the choice of basis set (see Section S.2.2 of the SM).

The fates of the photoexcited thiophenone molecules were explored further by running trajectory surface hopping (TSH) calculations from the photo-prepared $S_2$ state at the SA(4)-CASSCF(10/8) level of theory. These results are detailed in SM Section S2, and only the main features of the dynamics are highlighted here. As expected from inspection of the LIIC pathways, the initial $S_2$ population rapidly decays to the $S_1$ state and population appears almost immediately on the $S_0$ PES as shown in Fig. 3(b). All population is transferred to the $S_0$ state within 350 fs of UV excitation. Figure 3(c) displays a swarm of 46 TSH trajectories that mimic the relaxation dynamics of the thiophenone wavepacket and demonstrate that the ultrafast deactivation from $S_2$ to $S_1$ to $S_0$ is driven by a ballistic ring-opening process. The trajectories start to spread after ~50 fs; most remain ring-opened upon becoming $S_0^{\#}$ molecules, but some re-adopt a (vibrationally hot) cyclic configuration.

The conclusions from the TSH calculations match well with the experimental time-resolved photoelectron yields for the BE ranges selected to span the predicted $IP_{vert}$ values along the LIIC (Figs. 2(b) and 2(c)). The yield in the BE range corresponding to vertical ionization from the $S_2$ state (blue trace in fig. 2(b)) shows a narrow transient signal, the width of which is largely determined by the temporal instrument response function. With increasing $\Delta t$, this transient signal shifts to higher BE, broadens somewhat and gains a longer time tail (fig. 2(c)). Guided by the inset in Fig. 3(a), these observed changes are all consistent with the wavepacket evolving on the $S_2$ PES (sampled most cleanly by intensities at BE $\leq$ 5.5 eV), and subsequent non-adiabatic coupling with the $S_1$ state (which will be sampled most efficiently in the $6 \leq BE \leq 7$ eV range), and thence with the $S_0$ PES (which will start to be sampled at BE $\geq$ 7 eV). Ionization of $S_0^{\#}$ molecules accounts for the tails in the transients for the higher energy BE slices in Fig. 2(c); the build-up of $S_0^{\#}$ population (green crosses in Fig. 2(b)) plateaus at $\Delta t \geq 300$ fs. These comparisons serve to reinforce the interpretation developed from considering the LIIC pathways (Fig. 3(a)) that the experimentally observed increase in BE is a signature of the ultrafast decay of thiophenone ($S_2$) molecules to high $S_0^{\#}$ levels.

*Ground-state dynamics and reaction products.* One key feature of the present experimental study is that the response of UV-excited thiophenone molecules can be followed not just *en route* to, but also after reaching, the $S_0$ PES. Fig. 4(a) shows photoelectron spectra recorded at several pump-probe delays in the range $10 \leq \Delta t \leq 600$ ps. Most of the photoelectron intensity of interest at these large $\Delta t$ values lies in the range $8.0 \leq BE \leq 9.6$ eV, and the spectrum appears to consist of different contributions whose weights are $\Delta t$-dependent; the intensity of the feature at lower BE (peaking at BE ~8.8 eV) appears to increase relative to that of the feature peaking at BE ~9.3 eV. To simulate the ground state dynamics, the foregoing non-adiabatic molecular dynamics calculations using TSH in the lowest four electronic states were combined with *ab initio* molecular dynamics (AIMD, see Computational Details) on the $S_0$ PES. From an electronic structure perspective, this required switching from a SA-CASSCF description (used for the excited-state dynamics) to an unrestricted density functional theory (UDFT) picture using the PBE0 exchange/correlation functional; the SA-CASSCF active space employed for the TSH dynamics became unstable when the trajectories were prolonged on the $S_0$ PES, but AIMD with UDFT was found to offer a stable alternative and allowed long-time simulation of the $S_0^{\#}$ species (see, *e.g.* Ref. 31). Tests demonstrating the validity of this approach for the present system are reported in SM Section S2.5.

To initiate the $S_0^{\#}$ molecular dynamics after passage through the $S_1/S_0$ seam of intersection, the AIMD trajectories were launched from the nuclear coordinates and with the momenta given by the TSH trajectories upon reaching the $S_0$ state. Thus the present AIMD simulations are not *per se* in a ground-state thermal equilibrium, since the internal energy of the molecule at the start of the $S_0$-state dynamics calculation depends on the history of the TSH trajectory in the excited state, *i.e.* the approach allows description of non-statistical effects in the hot $S_0$-state dynamics. In total, 22 AIMD trajectories were propagated until $t = 2$ ps and, to explore the longer time dynamics, 10 of these were propagated further to $t = 100$ ps (see Computational Details and SM Section 2.10). Since each is a continuation of an excited-state trajectory, the starting configuration in each case involves a ring-opened or highly stretched molecule.

The AIMD simulations reveal formation of several different photoproducts within the earliest timescales of these dynamics. Ring closure (resulting in re-formation of 'hot' thiophenone molecules) is observed, as is the formation of the acyclic isomers 2-thioxoethylketene (P1), 2-(2-sulfanylethyl)ketene (P2), and 2-(2-thiiranyl)ketene (P3) (see Fig. 4(b) for structures). Interconversion between these isomers was observed in most trajectories within 2 ps (see *e.g.* SM Fig. S12). The histogram labelled 'other' in Fig. 4(b) includes all molecular geometries that could not be attributed to P1, P2, P3 or closed-ring thiophenone products. These rare 'other' geometries often correspond to transient configurations in the act of interconverting between the dominant photoproducts and are mostly observed within 500 fs of accessing the $S_0$-state PES. It is important to emphasize that these AIMD calculations (and the gas-phase experiments) involve isolated molecules. The potential energy acquired by thiophenone upon photoexcitation is converted, in part, to nuclear kinetic energy during the non-radiative decay to the $S_0$ state, but these are closed systems; no energy dissipation is possible and the resulting $S_0^{\#}$ species are highly energetic.

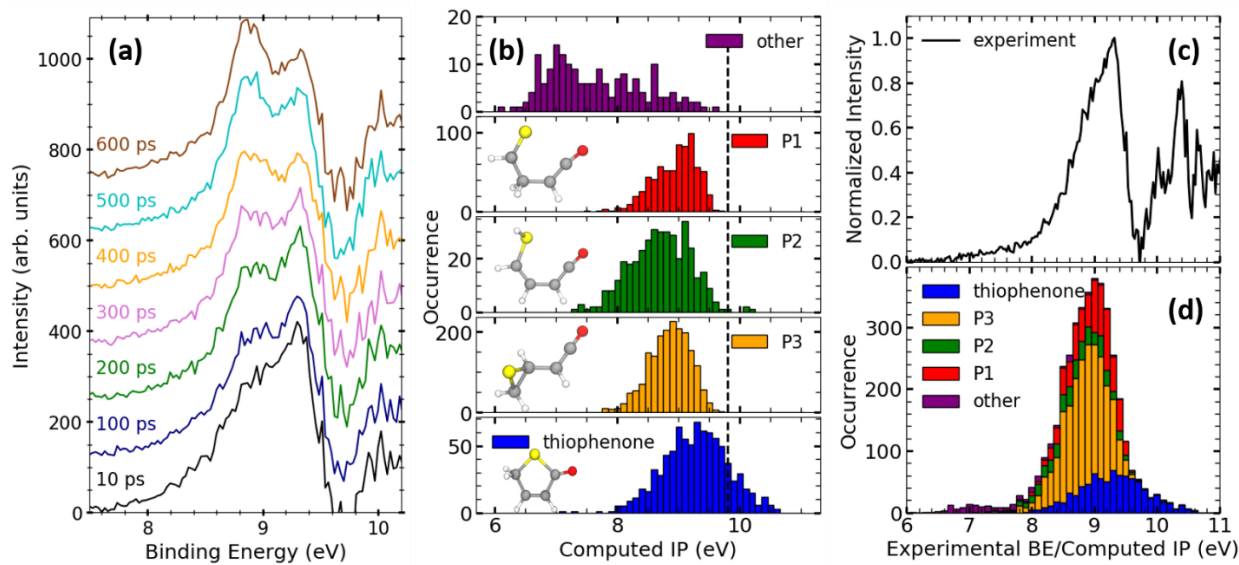

**Figure 4: Dynamics on the $S_0$ PES following photoexcitation and non-radiative decay.** (a) Photoelectron spectra for different pump-probe delays after subtraction of the signal from 'unpumped' thiophenone ($S_0$) molecules (see SM Fig. S3 for details). The spectra are offset vertically for better visibility. (b) Histograms showing the number of occurrences of the various $S_0 \rightarrow D_0$ $IP_{vert}$ values computed (at the MP2/cc-pVDZ-F12 level of theory) every 10 fs (for a total of 2 ps) along each AIMD trajectory on the $S_0$ PES, grouped by similarity to the molecular geometries identified as 2-thioxoethylketene (P1), 2-(2-sulfanylethyl)ketene (P2), 2-(2-thiiranyl)ketene (P3), and closed-ring thiophenone, along with a further small group labelled 'other' associated with internally hot molecules that are in the act of converting from one stable isomeric form to another at the time the trajectory was sampled (see text). The calculated $IP_{vert}$ for thiophenone at its optimized ground-state geometry is indicated by a dashed vertical black line. (c) Experimental (subtracted) photoelectron spectrum summed over the delay range $0.5 \leq \Delta t \leq 2$ ps; the contribution at BE >9.8 eV is due to ionization of $S_0^{\#}$ molecules to excited cationic states, which are not included in the present calculations (see also SM Figs. S2, S3, and S15). (d) Sum of the five distributions of $IP_{vert}$ values shown in panel (b).

Experimentally, the BEs of the $S_0^{\#}$ species formed via non-radiative de-excitation are concentrated in a narrow (~1 eV, full width at half maximum (FWHM)) band centered at ~9 eV, as shown in Figs. 4(a) and 4(c). Yet the AIMD simulations indicate that this ensemble of $S_0^{\#}$ species must contain a range of structures with high internal energy. Thus the 22 AIMD outputs were analyzed further. Specifically, the molecular geometry was extracted every 10 fs from each AIMD trajectory, yielding a pool of >4000 geometries. These were grouped by photoproduct (see SM Section S2.6) and, for each geometry, the $IP_{vert}$ value between the $S_0$ and $D_0$ states was calculated (using MP2-F12/cc-pVDZ-F12) to provide an estimate of the BE. As Fig. 4(b) shows, the histograms of the $IP_{vert}$ values for each photoproduct span a narrow range, and the ground state photoproducts P1, P2 and P3 display similar distributions of vertical IPs. This reflects the fact that, in each case, ionization involves removal of an electron from an orbital with a high degree of $n$(S) character. The distribution associated with re-formed closed-ring thiophenone species is centered at slightly higher $IP_{vert}$ values. For completeness, we note that the predicted distribution of $S_0 \rightarrow D_0$ $IP_{vert}$ values for thiophenone molecules without the extra kinetic energy induced by photoexcitation and subsequent relaxation (derived from AIMD simulations of thiophenone initiated with an internal energy equal to the zero-point energy only) is centered at yet higher $IP_{vert}$, is much narrower and, as SM Fig. S15 shows, is in very good accord with the lowest energy peak in the measured He I photoelectron spectrum[29].

Returning to the 'hot' $S_0^\#$ molecules, the 22 AIMD outputs predict a narrow overall distribution of $IP_{vert}$ values (Fig. 4(d)) that, as Fig. 4(c) shows, matches well with the TRPES data summed over the delay range $0.5 \leq \Delta t \leq 2$ ps. The 10 AIMD outputs propagated to $t = 100$ ps (see SM Fig. S17) provide additional insights into the longer time dynamics of these $S_0^\#$ molecules. Analysis of the (admittedly small number of) long time trajectories reveal (i) irreversible conversion of closed-ring to open-ring isomers and (ii), in several cases, unimolecular decay of the $S_0^\#$ species to CO + thioacrolein ($CH_2CHC(H)S$) products. The first $IP_{vert}$ of thioacrolein is 8.9 eV (see Ref. 32), a value reproduced computationally in the present work (see Section S2.10 in SM). Thus, recalling Fig. 4(b), we note that all closed-ring to open-ring transformations (including the fragmentation process) will cause a net transfer of $S_0^\#$ population to species with lower $IP_{vert}$ (*i.e.* lower BE) values – consistent with the experimental observations (Fig. 4(a)).

**Conclusions**

A previous transient infrared absorption spectroscopy study of the UV photoexcitation of thiophenone in solution (*i.e.* in an environment where any product vibrational excitation is rapidly dissipated through interaction with the solvent)[18] demonstrated the formation of acyclic photoproduct(s) with ketene structures and the recovery of (vibrationally cold) ground state thiophenone molecules. These earlier studies lacked the temporal resolution to probe the ring opening mechanism directly, and the only $S_0^\#$ dynamics amenable to investigation was vibrational relaxation via interaction with the solvent molecules. Such limitations are swept aside in the present study, wherein time-resolved XUV photoelectron spectroscopy studies of the isolated (gas phase) molecules at a seeded FEL, in combination with high-level *ab initio* theory, have allowed detailed visualization of the electronic and, particularly, the structural dynamics of this complex photo-induced ring-opening reaction, revealing the initial motion following photoexcitation, the non-adiabatic coupling to the $S_0$ PES as an open-ring biradical, and the subsequent isomerizations and eventual decay of the highly vibrationally excited $S_0$-state species. The match between theory and experiment spans both the excited-state decay rates and the more challenging athermal rearrangements that occur after non-adiabatic coupling to the $S_0$ state and onwards to photoproducts.

The use of sufficiently high-energy probe photons is key to tracking the full decay dynamics, *i.e.* the ultrafast evolution of the photoexcited wavepacket to the $S_0$ state and the nuclear dynamics of the resulting highly vibrationally excited $S_0$ molecules. The increase in BE observed immediately post-photoexcitation is a signature of the ultrafast decay of the nuclear wavepacket from the $S_2$ state, *via* the $S_1$ state, towards the electronic ground state, enabled by elongation and eventual scission of the S−CO bond. The evolving molecules couple to the $S_0$ PES with a range of geometries and nuclear momenta, which govern the subsequent athermal rearrangements of the $S_0^\#$ species. These vibrationally excited $S_0$ molecules are highly fluxional and can adopt at least three identified open-ring structures or re-form the parent thiophenone and have sufficient internal energy to dissociate (by loss of a CO moiety); the deduced ground state dynamics serve to bolster a recent prediction that isomerization of energized molecules prior to dissociation might well be the rule rather than the exception in many polyatomic unimolecular processes[33]. The distribution of vertical IPs computed from the AIMD trajectories on the $S_0$ PES reproduces the narrow spread of BEs observed experimentally and can be traced to the localized nature of the sulfur lone pair orbital that is the dominant contributor to the highest occupied molecular orbital in each species. Distinguishing the various open-ring products by valence-shell photoelectron spectroscopy is challenging given their very similar first IPs. Ultrafast X-ray or electron diffraction studies might be able to address such structural challenges if sufficient scattering signal can be obtained despite the low sample vapor pressure.

**Methods**

***Experimental:*** The experiment was performed at the low-density matter (LDM) beamline[34,35] at the FERMI free-electron laser facility[19]. The FEL was operated at a photon energy of 19.24 eV, corresponding to the fourth harmonic of the seed laser, with an estimated pulse duration of 80 fs (FWHM)[36]. The UV pump pulses with a center wavelength of 264.75 nm and 1.2 nm bandwidth were generated as the third harmonic of a Ti:Sapphire laser. Details relating to gas sample delivery, other laser pulse parameters (energies, durations and spot sizes) and tests to ensure that the reported effects scale linearly with pump and probe pulse parameters are reported in SM Section S1. A magnetic bottle type spectrometer[26,37] (see SM Fig. S1) was used to detect photoelectrons with high collection efficiency. A retardation voltage of 8 V was used to increase the resolution in the photoelectron range of interest (the approximate kinetic energy resolution is $\delta E/E \approx 0.03$, with $E$ being the final kinetic energy after retardation), cutting the photoelectron spectrum for binding energies of approximately 11 eV and above, as shown in SM Fig. S2.

***Data Acquisition and Analysis:*** Time-of-flight traces were recorded shot-by-shot while scanning the delay between the pump and probe pulses. The data shown in Fig. 1(a) consist of ~1650 shots per 50-fs delay bin. The single-shot spectra were normalized with respect to the FEL intensity (measured shot-to-shot[38]) and then averaged for each delay bin. The electron time-of-flight was converted into photoelectron kinetic energy by calibrating the spectrometer using photoelectrons from the single-photon single ionization of helium at multiple harmonics, *i.e.*, different photon energies, of the FEL. The photoelectron energies were then converted to binding energy by subtracting the photoelectron energy from the FEL photon energy.

***Computational Details:***
*Critical points and linear interpolation in internal coordinates*
Critical points of the thiophenone PESs – $S_0$ minimum, $S_1$ minima, $S_2/S_1$ and $S_1/S_0$ minimum energy conical intersections (MECIs) – were located using SA(4)-CASSCF(10/8)[39,40] and a 6-31G* basis set[41,42] as implemented in Molpro 2012[43]. Pathways connecting these different critical points of the PESs were produced by linear interpolations in internal coordinates (LIIC)[44]. Some of the advantages and limitations of LIICs are summarized in SM Section S2.2. Electronic energies for thiophenone were computed along the LIIC pathways at the SA(4)-CASSCF(10/8) and XMS(4)-CASPT2(10/8)[45,46] levels of theory using, in all cases, a 6-31G* basis set (see SM Figs. S7 and S8). The electronic energies for the thiophenone cation were also computed along the LIICs using SA(4)-CASSCF(9/8) and XMS(4)-CASPT2(9/8). All XMS-CASPT2 calculations were performed with the BAGEL software[47], employing the corresponding SA-CASSCF wavefunction from Molpro 2012 as a starting point. A level shift[48] of 0.3 $E_h$ was used in all XMS-CASPT2 calculations to prevent the appearance of intruder states. For details on the calculations of ionization potentials, including benchmarking studies justifying this choice of basis set, see SM Section S2.

*Trajectory surface hopping dynamics*

The excited-state dynamics of thiophenone following photoexcitation were simulated using the mixed quantum/classical dynamics trajectory surface hopping (TSH) method, employing the fewest-switches algorithm[49]. All details regarding these dynamics are provided in the SM Section S2.

*Ab initio molecular dynamics to t = 2 ps and t = 100 ps*

*Ab Initio* Molecular Dynamics (AIMD) calculations of the photoproducts formed during the TSH dynamics were conducted on the $S_0$-state PES using unrestricted DFT with the PBE0 exchange/correlation functional[50] and a 6-31G* basis set, employing the GPU-accelerated software TeraChem[51]. The initial conditions (nuclear coordinates and velocities) for each AIMD trajectory (22 in total, drawn randomly from the pool of TSH trajectories) were extracted from the TSH dynamics when the trajectory reached the $S_0$ state. At this initial point of configuration space, the SA-CASSCF wavefunction already exhibits a dominant closed-shell character (confirmed at the XMS-CASPT2 level of theory, see SM Section S2.5 for additional details). A small (0.1 fs) time step was used to ensure proper total energy conservation for all trajectories, and the length of each (constant total energy) trajectory was set such that the total TSH+AIMD dynamics extend to 2 ps. This strategy necessarily restricts the dynamics to the $S_0$ PES; the legitimacy of this procedure was validated by test trajectories on $S_0$, which show the energy separation between the ground and excited electronic states increasing rapidly upon leaving the region of the $S_1/S_0$ seam of intersection. To explore the long-time dynamics of the different photoproducts, 10 of the 22 trajectories were propagated further, to $t$ =100 ps, using the same methodology except for a slightly longer time step of 0.25 fs.

*Analysis of the 2-ps AIMD and vertical ionization energy distribution*

The 22 AIMD trajectories propagated until $t = 2$ ps were used to analyze the distribution of IP$_{vert}$ values for the $S_0^{\#}$ photoproducts. For each AIMD trajectory, molecular geometries were sampled every 10 fs, leading to a pool of >4000 $S_0$ molecular configurations. Each configuration was assigned to one of the possible photoproducts identified in Ref. 18 based on characteristic atomic connectivities determined by measuring bond lengths or angles (see SM Fig. S13). If such assignment was not possible, the configuration was given the label 'others'. These were often due to a transient configuration between two photoproducts. The IP$_{vert}$ of each configuration was then computed at the MP2-F12/cc-pVDZ-F12 level of theory (this level of theory was benchmarked against CCSD(T)-F12/cc-pVDZ-F12, see SM Section S2.8). The resulting distribution of $S_0 \rightarrow D_0$

IP$_{\text{vert}}$ values provides an approximation of the low-energy part of the experimental BE spectra. The same methodology, applied to ground-state dynamics of 'cold' thiophenone, successfully reproduces the first peak in the experimental He I photoelectron spectrum (see SM Fig. S15). All calculations were performed with Molpro 2012.

**Data availability**
The data that support the findings of this study are available from the corresponding authors upon reasonable request.

**Code availability**
The analysis codes used to generate the data presented in this study are available from the corresponding authors upon reasonable request.

**Acknowledgements**
S.P., J.T., and D.R. were supported by the National Science Foundation (NSF) grant PHYS-1753324. S.P. was also partially supported by the Chemical Sciences, Geosciences, and Biosciences Division, Office of Basic Energy Sciences, Office of Science, U.S. Department of Energy (DOE) under Grant No. DE-FG02-86ER13491. Travel to FERMI for S.P., D.M.P.H., R.M., J.T., and D.R. was supported by LaserLab Europe. M.N.R.A., C.S.H. and R.A.I. are grateful to the Engineering and Physical Sciences Research Council (EPSRC) for funding (EP/L005913/1), while L.M.I. acknowledges the EPSRC for a Doctoral Studentship (EP/R513039/1). C.S.H. als acknowledges funding from the Australian Research Council (ARC, DE200100549). M.N.R.A. is grateful to Prof W.-H. Fang (Beijing Normal University) for permission to share data from Ref. 28 prior to its publication. B.F.E.C. acknowledges funding from the European Union Horizon 2020 research and innovation programme under grant agreement No 803718 (SINDAM). D.M.P.H. was supported by the Science and Technology Facilities Council, UK. R.F. and R.J.S. acknowledge financial support from the Swedish Research Council, the Knut and Alice Wallenberg Foundation, Sweden, and from the Faculty of Natural Science of the University of Gothenburg. We thank the technical and scientific teams at FERMI for their hospitality and their support during the beamtime. We also acknowledge helpful discussions with Artem Rudenko during the preparation of the beamtime proposal and during the interpretation of the data and with Surjendu Bhattacharyya during the data analysis and interpretation.

**Author Contributions**
R.B., R.A.I., C.S.H., M.N.R.A. and D.R. conceived the experiment, the plans for which benefitted from further input from R.Fo., D.M.P.H. and A.R.. S.P., R.A.I., R.B., C.C., A.D., B.E., R.Fe., M.D.F., L.G., C.S.H., D.M.P.H., R.M., O.P., K.C.P., A.R., R.J.S., J.T., M.N.R.A. and D.R. conducted the experiment at the FERMI free-electron laser facility. The magnetic bottle spectrometer was provided and operated by R.Fe. and R.J.S.. C.C., M.D.F. and O.P. prepared and operated the beamline and the Low Density Matter (LDM) instrument. A.D. and L.G. prepared and operated the optical laser and the free-electron laser, respectively. L.M.I. and B.F.E.C. performed the ab-initio simulations with contributions from R.A.I.. S.P. and J.T. analyzed the experimental data with contributions from R.Fo., C.S.H., R.A.I., R.M. and A.R.. S.P., L.M.I., R.B.,

R.Fo., M.N.R.A., B.F.E.C. and D.R. interpreted the results and wrote the manuscript with input from all the authors.

**Competing interests**
The authors declare no competing interests.

**Additional information**
**Supplementary information** is available for this paper.

**Correspondence and requests for materials** should be addressed to D.R., B.F.E.C. or M.N.R.A.

**References**

1. Minitti, M.P. et al. Imaging Molecular Motion: Femtosecond X-Ray Scattering of the Ring Opening in 1,3- Cyclohexadiene. *Phys. Rev. Lett*. **114**, 255501 (2015).
2. Pullen, M. G. et al. Imaging an aligned polyatomic molecule with laser-induced electron diffraction. *Nature Communications* **6**, 7262 (2015).
3. Wolter, B. et al. Ultrafast electron diffraction imaging of bond breaking in di-ionized acetylene. *Science* **354**, 308–312 (2016).
4. Attar, A.R. et al. Femtosecond x-ray spectroscopy of an electrocyclic ring-opening reaction. *Science* **356**, 54–59 (2017).
5. Schuurman, M.S. & Stolow, A. Dynamics at Conical Intersections. *Annu. Rev. Phys. Chem*. **69**, 427–450 (2018).
6. Wolf, T.J.A. et al. The photochemical ring-opening of 1,3-cyclohexadiene imaged by ultrafast electron diffraction. *Nature Chemistry* **11**, 504–509 (2019).
7. Ruddock, J. M. et al. A deep UV trigger for ground-state ring-opening dynamics of 1,3-cyclohexadiene. *Science Advances* **5**, eaax6625 (2019).
8. Stankus, B. et al. Ultrafast X-ray scattering reveals vibrational coherence following Rydberg excitation. *Nature Chemistry* **11**, 716–721 (2019).
9. Browne, W. R. & Feringa, B. L. Light Switching of Molecules on Surfaces. *Annu. Rev. Phys. Chem*. **60**, 407–428 (2009).
10. Kumpulainen, T., Lang, B., Rosspeintner, A. & Vauthey, E. Ultrafast Elementary Photochemical Processes of Organic Molecules in Liquid Solution. *Chem. Rev*. **117**, 10826–10939 (2017).
11. Qi, J. et al. Light-driven transformable optical agent with adaptive functions for boosting cancer surgery outcomes. *Nature Communications* **9**, 1848 (2018).
12. Deb, S. & Weber, P. M. The Ultrafast Pathway of Photon-Induced Electrocyclic Ring-Opening Reactions: The Case of 1,3-Cyclohexadiene. *Annu. Rev. Phys. Chem*. **62**, 19–39 (2011).
13. Arruda, B.C. & Sension, R.J. Ultrafast polyene dynamics: the ring opening of 1,3-cyclohexadiene derivatives. *Phys. Chem. Chem. Phys*. **16**, 4439–4455 (2014).
14. Petrović, V. S. et al. Transient X-Ray Fragmentation: Probing a Prototypical Photoinduced Ring Opening. *Phys. Rev. Lett*. **108**, 253006 (2012).

15. Rudakov, F. & Weber, P. M. Ground State Recovery and Molecular Structure upon Ultrafast Transition through Conical Intersections in Cyclic Dienes. *Chem. Phys. Lett.* **470**, 187−190 (2009).
16. Adachi, S., Sato, M. & Suzuki, T. Direct Observation of Ground-State Product Formation in a 1,3-Cyclohexadiene Ring-Opening Reaction. *J. Phys. Chem. Lett.* **6**, 343–346 (2015).
17. Breda, S., Reva, I., Fausto, R. UV-induced unimolecular photochemistry of 2(5H)-furanone and 2(5H)-thiophenone isolated in low temperature inert matrices. *Vib. Spectrosc.* **50**, 57-67 (2009).
18. Murdock, D. et al. Transient UV pump–IR probe investigation of heterocyclic ring-opening dynamics in the solution phase: the role played by nσ* states in the photoinduced reactions of thiophenone and furanone. *Phys. Chem. Chem. Phys*. **16**, 21271–21279 (2014).
19. Allaria, E. et al. Highly coherent and stable pulses from the FERMI seeded free-electron laser in the extreme ultraviolet. *Nature Photonics* **6**, 699–704 (2012).
20. Stolow, A., Bragg, A. E. & Neumark, D. M. Femtosecond Time-Resolved Photoelectron Spectroscopy. *Chem. Rev*. **104**, 1719–1757 (2004).
21. Suzuki, T. Time-resolved photoelectron spectroscopy of non-adiabatic electronic dynamics in gas and liquid phases. *Inter. Rev. Phys. Chem*. **31**, 265–318 (2012).
22. Iikubo, R., Sekikawa, T., Harabuchib, Y. & Taketsugubet, T. Structural dynamics of photochemical reactions probed by time-resolved photoelectron spectroscopy using high harmonic pulses. *Faraday Discussions* **194**, 147–160 (2016).
23. Nishitani, J., West, C.W., Higashimura, C. & Suzuki, T. Time-resolved photoelectron spectroscopy of polyatomic molecules using 42-nm vacuum ultraviolet laser based on high harmonics generation. *Chemical Physics Letters* **684**, 397–401 (2017).
24. Smith, A.D. et al. Mapping the Complete Reaction Path of a Complex Photochemical Reaction. *Phys. Rev. Lett.* **120**, 183003 (2018).
25. von Conta, A. et al. Conical-intersection dynamics and ground-state chemistry probed by extreme-ultraviolet time-resolved photoelectron spectroscopy. *Nature Communications* **9**, 3162 (2018).
26. Squibb, R.J. et al. Acetylacetone photodynamics at a seeded free-electron laser. *Nature Communications* **9**, 63 (2018).
27. Gorobtsov, O. Y., et al., Seeded X-ray free-electron laser generating radiation with laser statistical properties, *Nature Communications* **9,** 4498 (2018).
28. Xie, B.-B. & Fang, W.-H. Combined Quantum Trajectory Mean-Field and Molecular Mechanical (QTMF/MM) Nonadiabatic Dynamics Simulations on the Photoinduced Ring-Opening Reaction of 2(5H)-Thiophenone. *ChemPhotoChem* **3**, 897–906 (2019).
29. Chin, W.S. et al. He I and He II photoelectron spectra of thiophenones. *J. Electron Spectroscopy Related Phenomena* **88–91**, 97–101 (1998).
30. Tao, H. et al. Ultrafast internal conversion in ethylene. I. The excited state lifetime. *J. Chem. Phys*. **134**, 244306 (2011).

31. Mignolet, B., Curchod, B. F. E. & Martinez, T. J. Rich Athermal Ground-State Chemistry Triggered by Dynamics through a Conical Intersection. *Angew. Chem., Int. Ed.* **55**, 14993−14996 (2016).
32. Bock, H., Mohmand, S., Hirabayashi, T. & Semkow, A. Gas-phase reactions. 29. Thioacrolein. *J. Am. Chem. Soc.* **104**, 312-313 (1982).
33. Yang, C.-S., Bhattacharyya, S., Liu, L., Fang, W.-h. & Liu, K. Real-time tracking of the entangled pathways in the multichannel photodissociation of acetaldehyde. *Chem. Sci.* DOI: 10.1039/d0sc00063a
34. Svetina, C. et al. The Low Density Matter (LDM) beamline at FERMI: optical layout and first commissioning. *J. Synchrotron Rad.* **22**, 538–543 (2015).
35. Lyamayev, V. et al. A modular end-station for atomic, molecular, and cluster science at the low density matter beamline of FERMI@Elettra. *J. Phys. B: At. Mol. Opt. Phys.* **46**, 164007 (2013).
36. Finetti, P. et al. Pulse Duration of Seeded Free-Electron Lasers. *Phys. Rev. X* **7**, 021043 (2017).
37. Eland, J. H. D. et al. Complete two-electron spectra in double photoionization: the rare gases Ar, Kr, and Xe. *Phys. Rev. Lett.* **90**, 053003 (2003).
38. Zangrando, M. et al. Recent results of PADReS, the Photon Analysis Delivery and REduction System, from the FERMI FEL commissioning and user operations. *J. Synchrotron Rad.* **22**, 565–570 (2015).
39. Werner, H.-J. Matrix-formulated direct multiconfiguration self-consistent field and multiconfiguration reference configuration-interaction methods. *Advances in Chemical Physics: Ab Initio Methods in Quantum Chemistry Part 2* **69**, 1–62 (1987).
40. Roos, B. O. The complete active space self-consistent field method and its applications in electronic structure calculations. *Advances in Chemical Physics: Ab Initio Methods in Quantum Chemistry Part 2* **69**, 399–445 (1987).
41. Hehre, W. J., Ditchfield, R. & Pople, J. A. Self-Consistent Molecular Orbital Methods. XII. Further extensions of Gaussian-type basis sets for use in molecular-orbital studies of organic-molecules. *J. Chem. Phys.* **56**, 2257- 2261(1972).
42. Hariharan, P. C. & Pople, J. A. Influence of polarization functions on molecular-orbital hydrogenation energies. *Theor. Chem. Acc.* **28**, 213–22 (1973).
43. Werner, H.-J. et al. MOLPRO, version 2012.1, a package of *ab initio* programs (http://www.molpro.net) (2012).
44. Hudock, H. R. et al. Ab initio molecular dynamics and time-resolved photoelectron spectroscopy of electronically excited uracil and thymine. *The Journal of Physical Chemistry A* **111**, 8500–8508 (2007).
45. Shiozaki, T., Győrffyet W., Celani, C. & Werner, H.-J. Communication: Extended Multi-State Complete Active Space Second-Order Perturbation Theory: Energy and Nuclear Gradients. *J. Chem. Phys.* **135**, 081106 (2011).
46. Granovsky, A. Extended Multi-Configuration Quasi- Degenerate Perturbation Theory: The New Approach to Multi-State Multi-Reference Perturbation Theory. *J. Chem. Phys.* **134**, 214113 (2011).

47. BAGEL Brilliantly Advanced General Electronic-structure Library, http://www.nubakery.org under the GNU General Public License
48. Ghigo, G., Roos, B. O & Malmqvist, P.A. A modified definition of the zeroth-order Hamiltonian in multiconfigurational perturbation theory (CASPT2). *Chem. Phys. Lett.* **396**, 142-149 (2004).
49. Tully, J.C. Molecular dynamics with electronic transitions. *J. Chem. Phys.* **93**, 1061–1071 (1990).
50. Adamo, C. & Barone, V. Toward reliable density functional methods without adjustable parameters: The PBE0 model. *J. Chem. Phys*. **110**, 6158–69 (1999).
51. Ufimtsev, I.S. & Martinez, T. J. Quantum Chemistry on Graphical Processing Units. 3. Analytical Energy Gradients, Geometry Optimization, and First Principles Molecular Dynamics. *J. Chem. Theory Comput*. **5**, 2619–2628 (2009).

## Supplementary Material

*S1. Experimental details*

2(5H)-thiophenone (Sigma Aldrich, 98%) seeded in a helium carrier gas at 2 bar backing pressure was introduced into the vacuum chamber through a pulsed Even-Lavie valve heated to 60 ºC. The FEL pulses were focused to a spot of 40 x 50 µm$^2$ (FWHM) and had an average pulse energy of 18 µJ (measured before the beamline optics). The calculated transmission of the transport optics at this wavelength is 0.27 [*Svetina2015*], implying an average pulse energy of 4.9 µJ at the sample. The UV pump pulses (center wavelength: 264.75 nm, bandwidth: 1.2 nm) were generated as the third harmonic of a Ti:Sapphire laser and had an average pulse energy of 25 µJ, focused to a diameter of ~200 µm (FWHM). To investigate the dependence of the observed effects on the pulse energy of the pump and probe pulses, delay scans were taken at several lower and higher pulse energies (see SM Fig. S4). The observed spectral components and time scales were found to be independent of the pulse energies.

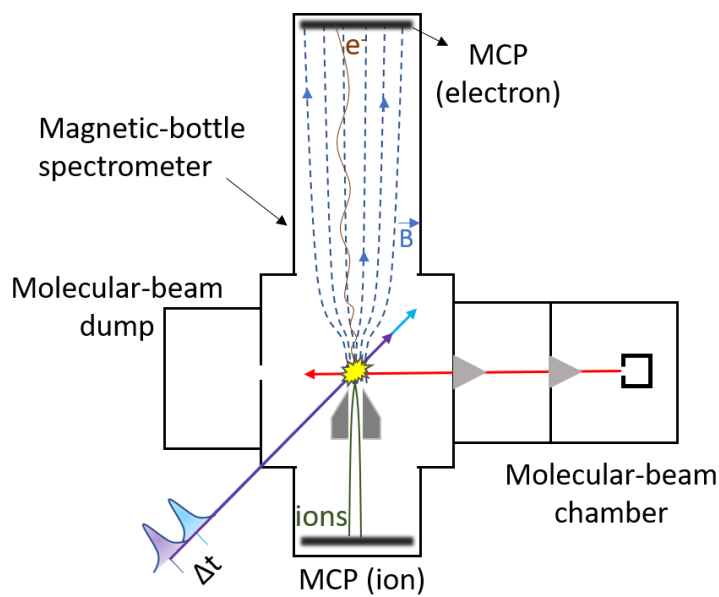

**Figure S1: Schematic of experimental setup.** The FEL and UV pulses intersect a cold beam of target molecules seeded in helium carrier gas that is introduced into the vacuum chamber by supersonic expansion through a pulsed valve. Photoelectrons created by the interaction of the FEL pulses with the target are guided onto an MCP detector by the electric and magnetic (**B**) field of the magnetic bottle spectrometer. Photoions, although not discussed further in this work, are extracted through a hole in the magnet of the magnetic bottle and are detected on a second MCP detector opposite to the electron detector.

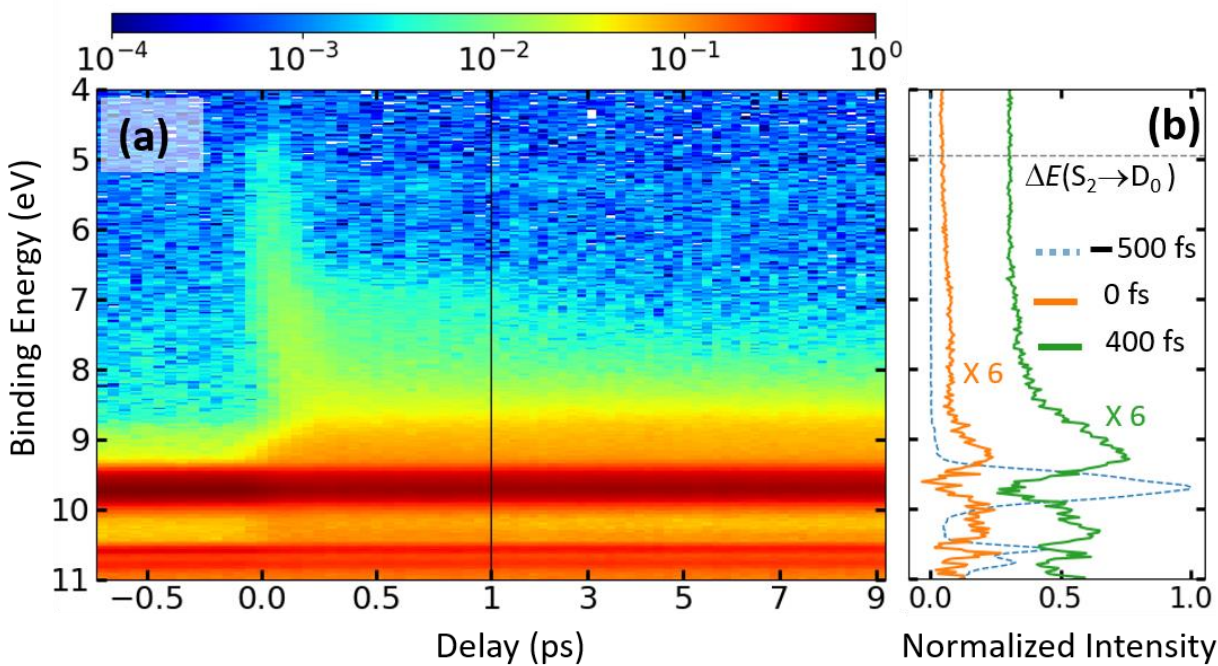

**Figure S2: Measured photoelectron yield as a function of binding energy and pump-probe delay.** Note the different scaling of the delay axis below and above $\Delta t = 1.0$ ps, and the larger binding energy range and different color scale than used in Fig. 2(a). The strong contributions at BE ~9.7 and BE ~10.6 eV correspond to ionization to the $D_0/D_1$ and the $D_2$ states of the cation, respectively [*Chin1998*]. Photoelectrons with BE >11 eV were not detected because a retardation voltage of 8 V was used to obtain better resolution for the electrons with lower binding energies. (b) Photoelectron spectra for the FEL pulse preceding the UV pulse ($\Delta t = -500$ fs, dashed blue line) and at two delays where the UV pulse precedes the FEL pulse (orange: $\Delta t = 0$; green: $\Delta t = 400$ fs; scaled by a factor of 6 and offset by, respectively, 0.04 and 0.3 for better visibility). For the latter two, the contributions from 'unpumped' molecules have been subtracted (see Fig. S3), and the data are integrated over a range of ±40 fs around the indicated $\Delta t$ value. The ionization potential for vertical ionization (in the ground-state equilibrium geometry) from the $S_2$ excited state to the $D_0$ state of the cation (taken from [*Chin1998*]) is shown as a dashed line.

**Table S1: Least-square fits of the delay-dependent photoelectron yields.** Fit parameters obtained from least-square fits to the lineouts of the photoelectron yield in selected binding energy ranges shown in Fig. 2(b). The data are fitted by a Gaussian (G(x)), the convolution C(x) = (f*h)(x) of a Gaussian (h(x)) with an exponential decay (f(x)); or by a cumulative distribution function (CDF(x)), as appropriate and as defined below.

| Energy Range | Fit Function | $x_0$/fs | $\sigma$/fs | $\tau$/fs |
|---|---|---|---|---|
| (4.7 – 5.3) eV | Gaussian | 4 ± 5 | 76 ± 6 | - |
| (8.3 – 9.1) eV | Gaussian * Exponential | -2 ± 8 | 75 (fixed) | 72 ± 9 |
| (9.5 – 9.9) eV | CDF | -15 ± 5 | 72 ± 8 | - |

$$G(x) = Ae^{\frac{-(x-x_0)^2}{2\sigma^2}} + y_0$$

$$CDF(x) = \frac{A}{2}\left(1 + \text{erf}\left(\frac{x - x_0}{\sqrt{2\sigma^2}}\right)\right) + B$$

Convolution of Gaussian $h(x) = e^{-\frac{x^2}{2\sigma^2}}$ and

exponential (for x > x0) $f(x) = \text{Heaviside}(x - x0)\left\{A\left(1 - e^{\frac{-(x-x0)}{\tau}}\right) + B\right\}$:

$$C(x) = \sigma e^{-\frac{x}{\tau}}\sqrt{\frac{\pi}{2}}\left((A+B)e^{\frac{x}{\tau}}\left(1 + \text{Erf}\left[\frac{x-x0}{\sigma\sqrt{2}}\right]\right) - Ae^{\frac{2\tau x0 + \sigma^2}{2\tau^2}}\text{Erfc}\left[\frac{(-\tau(x-x0)+\sigma^2)}{\sigma\sqrt{2}\tau}\right]\right)$$

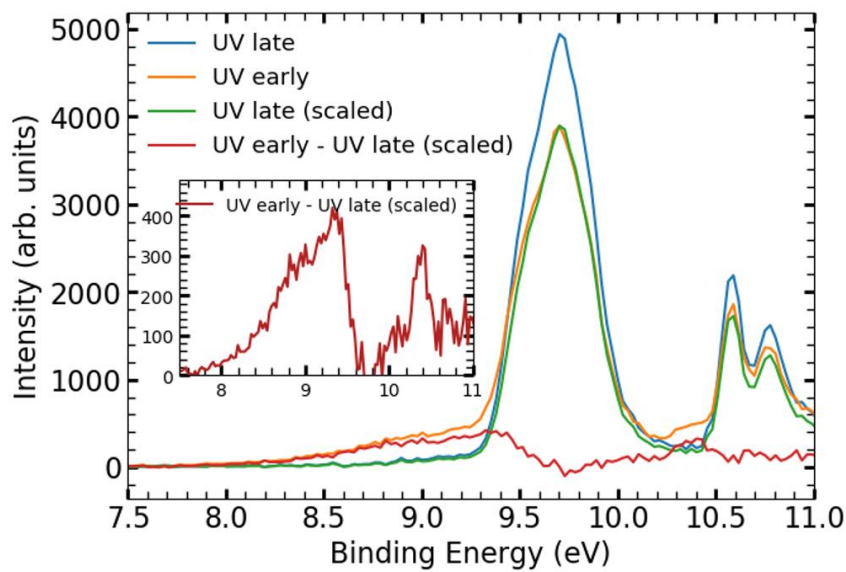

**Figure S3: Subtraction of signal from 'unpumped' molecules.** To isolate the time-dependent signal stemming from photo-excited molecules, the photoelectron spectrum measured at negative $\Delta t$ (*i.e.* FEL pulse arriving before the UV pulse, blue curve, 'UV late') is used in order to subtract the signal from unpumped molecules at positive delays (orange curve, 'UV early'). For delays $\Delta t >0.2$ ps, the fraction of unpumped molecules is constant, and a constant scaling factor is chosen. However, in the temporal overlap region, the subtraction procedure needs to recognize the gradual decrease of this signal caused by the increasing ground-state depletion. To determine the appropriate scaling factor for a given delay, the 'UV late' spectrum is scaled such that it has the same photoelectron yield in the binding energy range between 9.5 and 9.9 eV as the spectrum at the given delay. This procedure is repeated for all delay points in the overlap region, and a cumulative distribution function (CDF) is fitted to obtain a smooth scaling function that is used to generate the subtracted spectra at all delays. The red line (repeated in the inset on an expanded vertical scale) shows an example of a subtracted spectrum at $\Delta t = 1.28$ ps). The contribution at BE >9.8 eV in the subtracted spectrum is due to ionization of $S_0^{\#}$ molecules to excited cationic states (see also Fig. S15), which are not included in the calculations presented in this work.

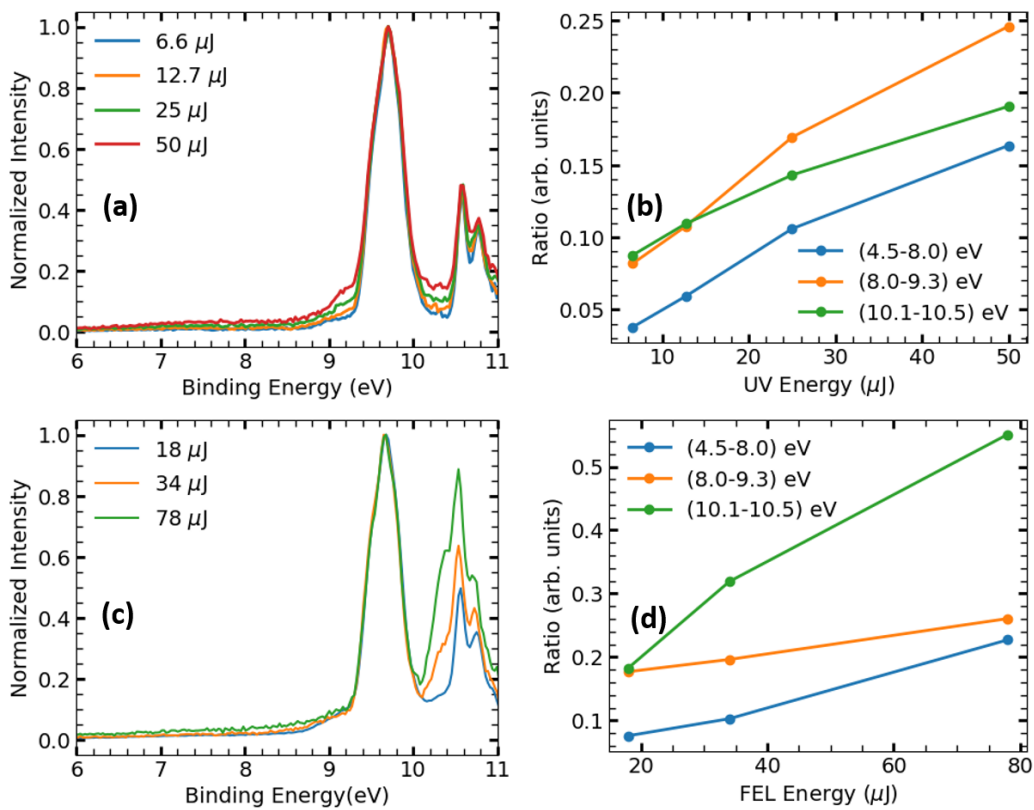

**Figure S4: Pulse-energy dependence of the pump-probe signal.** Dependence of the photoelectron spectra on the UV (a, b) and FEL (c, d) pulse energies. The latter are measured *before* the beamline optics and do not include the beamline transmission [*Svetina2015*]. The spectra in (a) and (c) are normalized with respect to the photoelectron signal intensity from unpumped molecules (BE ~9.6 eV). Panels (b) and (d) show the ratio of the photoelectron yield in the specified binding energy ranges with respect to the yield from unpumped molecules (BE 9.5-9.9 eV). The data shown in the manuscript were recorded at FEL and UV pulse energies of 19 μJ and 25 μJ, respectively.

### S2. Computational details

This section starts by providing additional details for the calculations of ionization potentials and surface hopping dynamics. We also provide additional benchmarks and supporting discussions relating to the calculations reported in the main text.

*Ionization potentials for potential photoproducts and benchmarking:* Ground-state geometry optimizations for thiophenone and its proposed photoproducts [*Murdock2014*] were conducted *in vacuo* at the MP2/6-311+G** level of theory, and the nature of all located stationary points was verified by harmonic vibrational frequency calculations. Based on these geometries, single-point electronic energy calculations were performed using MP2-F12/cc-pVDZ-F12 and CCSD(T)-F12/cc-pVDZ-F12 [*Kong2012, Hattig2012*]. Both methods are in good agreement for all the computed ionization potentials of each molecule (see SM Section S2.8). All calculations were performed with Molpro 2012.

*Trajectory surface hopping dynamics:* The excited-state dynamics of thiophenone following photoexcitation were simulated using the mixed quantum/classical dynamics method trajectory surface hopping (TSH), employing the fewest-switches algorithm proposed by Tully [*Tully1990*]. The calculations were performed with the SHARC program package (v2.0) [*Mai2018*]. 46 initial conditions for the TSH dynamics were sampled stochastically from a Wigner distribution for uncoupled harmonic oscillators constructed from a frequency calculation at the ground-state optimized geometry of thiophenone. All trajectories were initiated in the bright $S_2$ state of thiophenone (see SM Section S2.4 for more information). The TSH dynamics employed a time step of 0.5 fs and SA(4)-CASSCF(10/8) for the electronic structure (benchmarking of this method is discussed more fully in SM Section S2.2) using Molpro 2012. The energy decoherence correction scheme by Granucci and Persico [*Granucci2007*] was applied to the electronic coefficients with the default constant of 0.1 hartree. Strict total energy conservation was ensured for each trajectory during the excited-state dynamics. However, the active space showed instabilities within a few tens of femtoseconds in the $S_0$ state. Such instabilities did not constitute an issue as the TSH dynamics were sufficiently stable to provide initial conditions for the subsequent ground-state AIMD calculations.

### *S2.1. Characterization of the excited states*

The electronic character of the first two excited singlet states of thiophenone at the Franck-Condon geometry was determined by natural difference orbitals (NDO), computed with the TheoDORE program (v 2.0) [*Plasser2014*].

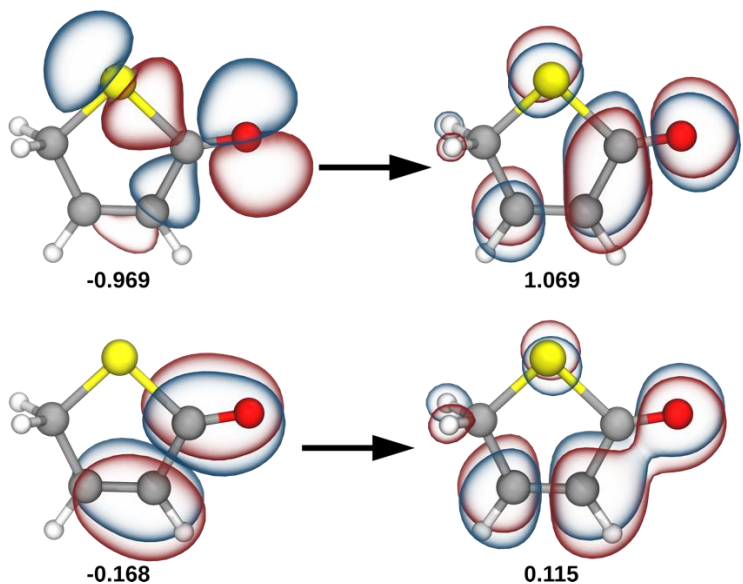

**Figure S5: Electronic character of the $S_1$ state.** Pairs of NDOs contributing to the $S_1$ state at the Franck-Condon geometry, with their respective eigenvalues.

Figures S5 and S6 show the NDOs (the eigenvectors of the difference density with respect to the ground electronic state) for the $S_1$ and $S_2$ states, respectively, with their corresponding eigenvalues; where negative (positive) eigenvalues describe the detachment (attachment) process. For detailed information about NDOs we refer to [*Plasser2014*]. The first pair of NDOs in Fig. S5 shows that the $S_1$ state is dominated by a $n(O)/\pi^*$ transition, with a small contribution of a $\pi/\pi^*$ transition. The NDOs for the $S_2$ state (Fig. S6) encourage assignment as a $n(S)/\pi^*$ transition, but there is also a smaller but significant contribution from the $n(O)/\sigma^*$ transition (highlighting the dissociative character of this state) and a minor contribution from a $\pi/\pi^*$ transition (similar to the small contribution in $S_1$).

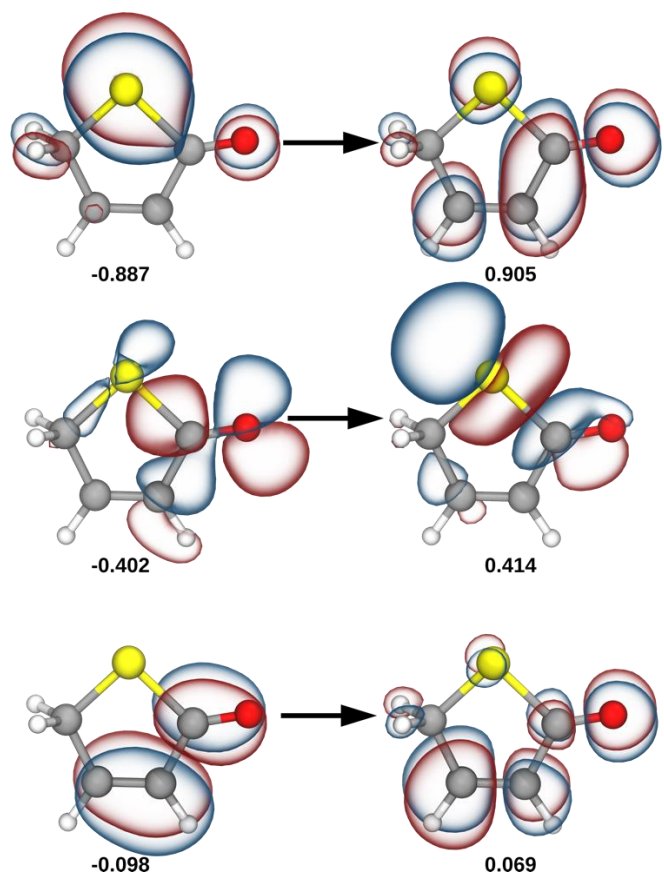

**Figure S6: Electronic character of the $S_2$ state.** Pairs of NDOs contributing to the $S_2$ state at the Franck-Condon geometry, with their respective eigenvalues.

### *S2.2. LIIC pathways – validation of the electronic structure methods*

LIIC pathways allow determination of the most straightforward path from geometry A to geometry B by interpolating a series of intermediate geometries (ten in the present work), using internal (not Cartesian) coordinates. It is important to recognize that *no reoptimization* is performed along these pathways. As a result, LIIC pathways should not be compared directly with minimum energy paths, as the barriers observed in LIICs are upper estimates of the barriers that would be returned by locating the true transition states.

The LIIC pathways presented in the main text were derived from critical geometries optimized at the SA(4)-CASSCF(10/8) level of theory, but their energies were refined with XMS(4)-CASPT2(10/8). Figure S7 (right half) shows the same LIIC pathway calculated at the SA(4)-CASSCF(10/8) level. The overall shape of this pathway is in excellent agreement with that obtained using XMS(4)-CASPT2(10/8) (which is reproduced again in the right-hand part of Fig. S8 below), validating the use of SA-CASSCF for the nonadiabatic molecular dynamics. We note that the SA-CASSCF $S_2/S_1$ MECI point is not perfectly degenerate at the XMS-CASPT2 level of theory, while the $S_1/S_0$ MECI point is. The variations in ionization energy along the LIIC pathway as given by SA-CASSCF (shown at the top of Fig. S7) are also in excellent agreement with that returned by XMS-CASPT2.

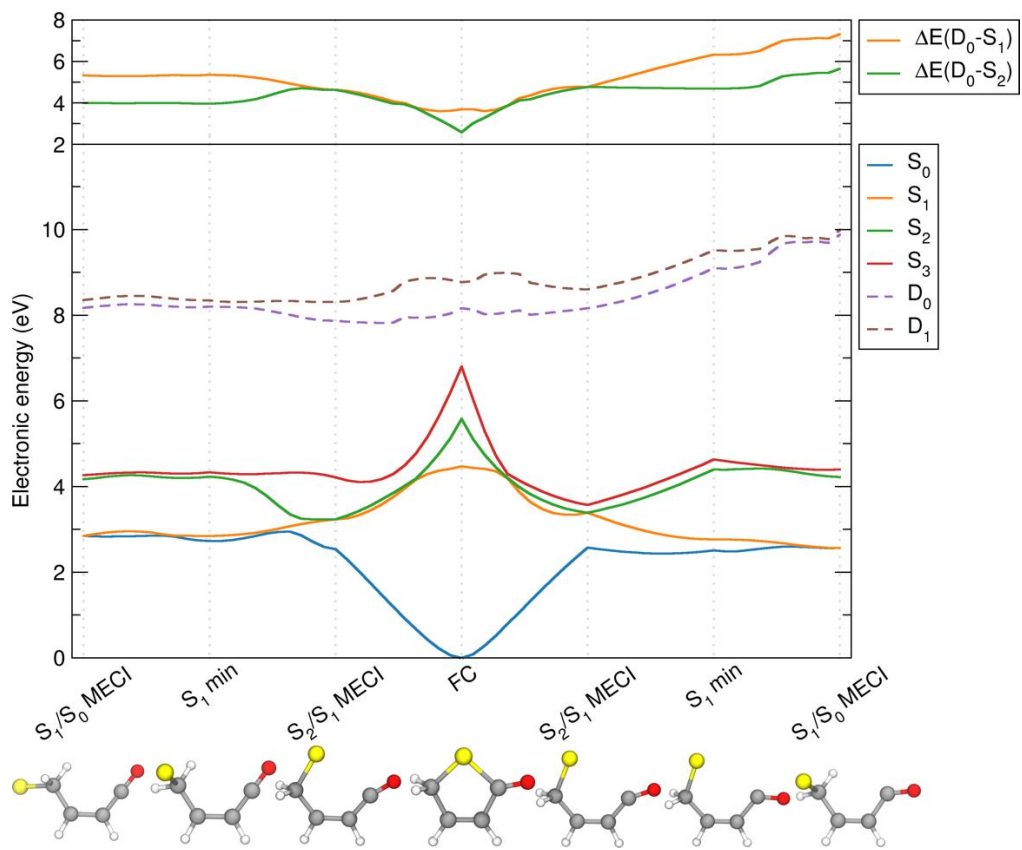

**Figure S7: LIIC pathways.** LIIC pathways computed at the SA(4)-CASSCF(10/8) (neutral thiophenone, solid lines) and SA(4)-CASSCF(9/8) (thiophenone cation, dashed lines) levels of theory. The LIIC pathway to the right of the FC region is that discussed in the main text, while that shown to the left is discussed below. Molecular geometries of the critical points located at the SA(4)-CASSCF(10/8) level of theory are represented beneath the figure.

The left part of Fig. S7 presents an alternative LIIC pathway connecting the FC point on the $S_2$ state to the $S_0$ state. This path differs in that the critical geometries located for the $S_1$ minimum and the $S_1/S_0$ MECI exhibit an out-of-plane configuration and a nonlinear C=C=O moiety (depicted at the appropriate points at the bottom of Fig. S7). The overall connectivity between $S_2$, $S_1$, and $S_0$ is nevertheless very similar to that observed for the LIIC pathway shown on the right: direct decay from the FC point on $S_2$ through conical intersections toward $S_0$. Importantly, the majority (76%) of the last $S_1$-to-$S_0$ hops observed during the TSH dynamics (discussed below) took place for molecular configurations similar to the $S_1/S_0$ MECI of the LIIC pathway on the right (*i.e.* the one presented in the main text). Figure S9 shows this comparison. We note the excellent agreement between the overall shape of this second LIIC pathway as determined by SA-CASSCF and XMS-CASPT2 methods (compare Figs. S7 and S8).

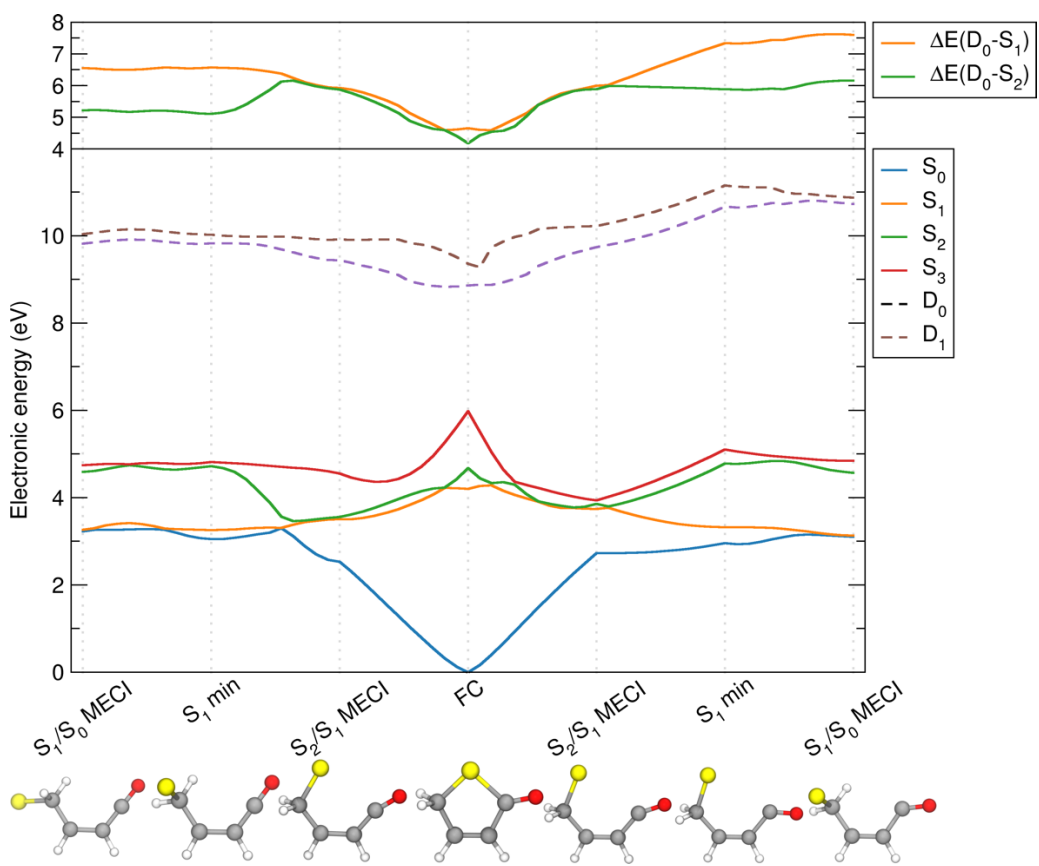

**Figure S8: LIIC pathways with refined energies.** LIICs pathways presented in Fig. S7 with energies refined at the XMS(4)-CASPT2(10/8) (neutral thiophenone, solid lines) and XMS(4)-CASPT2(9/8) (thiophenone cation, dashed lines) levels of theory. Molecular geometries of the critical points located at the SA(4)-CASSCF(10/8) level of theory are represented beneath the figure.

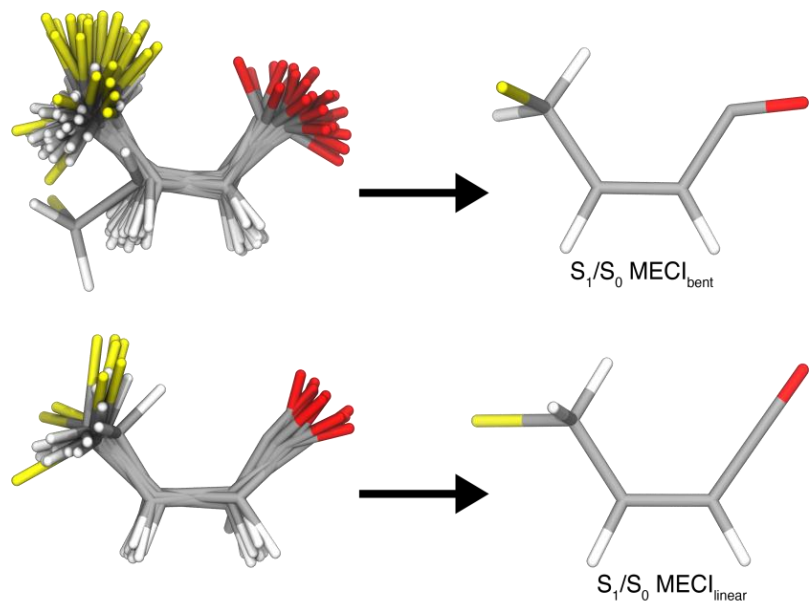

**Figure S9: Geometries at the end of TSH calculations.** Superimposed last $S_1/S_0$ hopping geometries from the TSH dynamics (shown on the left), grouped into two families according to whether the C=C=O angle corresponds to the bent MECI (76% of the hopping geometries, upper panel) or linear MECI (24% of the hopping geometries, lower panel).

Table S2 illustrates the influence of the choice of basis set on the XMS-CASPT2 calculations for the Franck-Condon geometry of thiophenone. At the SA-CASSCF level of theory, use of the larger cc-pVTZ basis set has negligible difference on the excitation energies for either neutral or charged thiophenone, and only a small, essentially rigid shift of all transitions energies is observed within XMS-CASPT2. Interestingly, the energy gap between the neutral and cationic states is increased when increasing the size of the basis set: for example, XMS-CASPT2/cc-pVTZ returns a IP$_{vert}$ value for the $S_0 \rightarrow D_0$ ionization of 9.38 eV, *cf.* 8.94 eV for XMS-CASPT2/6-31G*. This accounts for the difference between the computed and measured BEs highlighted in the main text (inset of Fig. 3(a)). Nevertheless, we have used the smaller basis set to explore the PESs of thiophenone and for the TSH dynamics as a compromise between accuracy and efficiency and because the benchmarking assures us that this causes only a rigid energy shift.

**Table S2: Illustration of the effect of the basis set on the electronic energies of thiophenone (at the Franck-Condon geometry).** Energies are given in eV, with respect to the corresponding $S_0$ electronic energy.

| Neutral | SA(4)-CASSCF(10/8) 6-31G* | SA4-CASSCF(10/8) cc-pVTZ | XMS(4)-CASPT2(10/8) 6-31G* | XMS(4)-CASPT2(10/8) cc-pVTZ |
|---|---|---|---|---|
| $S_0$ | 0 | 0 | 0 | 0 |
| $S_1$ | 4.477 | 4.475 | 4.199 | 4.063 |
| $S_2$ | 5.577 | 5.505 | 4.676 | 4.475 |
| $S_3$ | 6.800 | 6.758 | 5.985 | 5.747 |
| **Cation** | SA(4)-CASSCF(9/8) 6-31G* | SA4-CASSCF(9/8) cc-pVTZ | XMS(4)-CASPT2(9/8) 6-31G* | XMS(4)-CASPT2(9/8) cc-pVTZ |
| $D_0$ | 8.293 | 8.276 | 8.940 | 9.377 |
| $D_1$ | 8.827 | 8.765 | 9.407 | 9.757 |

### *S2.3 Selection rules for ionization from $S_2$ to $D_0$ and $D_1$*

A key focus of the present study is the energetic separation between the neutral and cationic states of thiophenone along the LIIC pathway. Here it may be useful to introduce briefly an electronic argument for focusing the discussion on ionization to $D_0$ rather than to the $D_1$ state. The $S_2$ state has dominant $n(S)/\pi^*$ character in the Franck-Condon region (see Fig. S6). Importantly, the S atom lone pair remains the dominant donating orbital along the entire decay pathway, from the Franck-Condon point on $S_2$, to the $S_2/S_1$ MECI, and during the relaxation on $S_1$ towards $S_1$ min, eventually reaching the $S_1/S_0$ MECI. Similarly, the $D_0$ state is always formed by loss of an electron from the S atom lone pair along this LIIC pathway, whereas the $D_1$ state is initially (*i.e.* in the Franck-Condon region) characterized by removal of an electron from the O lone pair, before it too gains a larger contribution from the $n(S)$ orbital towards the end of the LIIC (when the states get closer in energy). Simple ionization rules [*Arbelo-González2016*] suggest that ionization should be allowed from $S_2$ to $D_0$ along the entire LIIC pathway, as the two states differ by just a single spin-orbital (the accepting orbital in the excitation of the neutral). However, the same logic would imply that ionization from $S_2$ to $D_1$ is disfavored, as it necessitates (at least in the first part of the LIIC) a change of more than one spin-orbital in the ionization step.

### *S2.4 Trajectory Surface Hopping dynamics of thiophenone*

The electronic-state population trace obtained with TSH for 46 independent trajectories is shown in Fig. 3 of the main text. The trajectories were all initialized on the $S_2$ state, since the computed vertical transitions (at the SA(4)-CASSCF(10/8) level of theory) of 100 Wigner sampled structures indicate that this state is by far the 'brightest' when exciting from the $S_0$ state (see Fig. S10) and lies at the appropriate energy for the chosen pump wavelength (when the transition energy is computed at the XMS-CASPT2 level of theory, see main text).

The populations depicted in the main text are computed as the fraction of trajectories evolving in a given electronic state at a given time. We note that the displayed population traces match well with those computed from the squares of the TSH electronic coefficients, averaged over all trajectories – indicating an internal consistency of the TSH algorithm. In accord with the LIIC pathways presented above, the nuclear wavepacket created on $S_2$ exhibits an ultrafast (<100 fs) decay towards the lower electronic states. The growth of the $S_1$ population, if fitted by a single exponential, is characterized by a waiting period of 19 fs and a growing time of 83 fs.

To shed light on the possible involvement of triplet states, we also investigated the effects of including two triplet states into the TSH dynamics. The TSH trajectories were initialized on $S_2$, as in the singlet-only TSH dynamics. The presence of triplet states was found not to influence the initial ring opening upon excitation. After internal conversion to the $S_1$ state, these triplet states become almost degenerate with the $S_1$ and $S_0$ states. By opening new decay channels, the presence of the triplet states could result in a slower decay to the ground state. In these test calculations, the population transfer to the triplet states was observed to be minor (<20% of the total electronic population). However, the instability encountered during the dynamics upon addition of triplet states in the SA-CASSCF calculations prevents us from drawing quantitative conclusions and these observations need to be regarded with due caution.

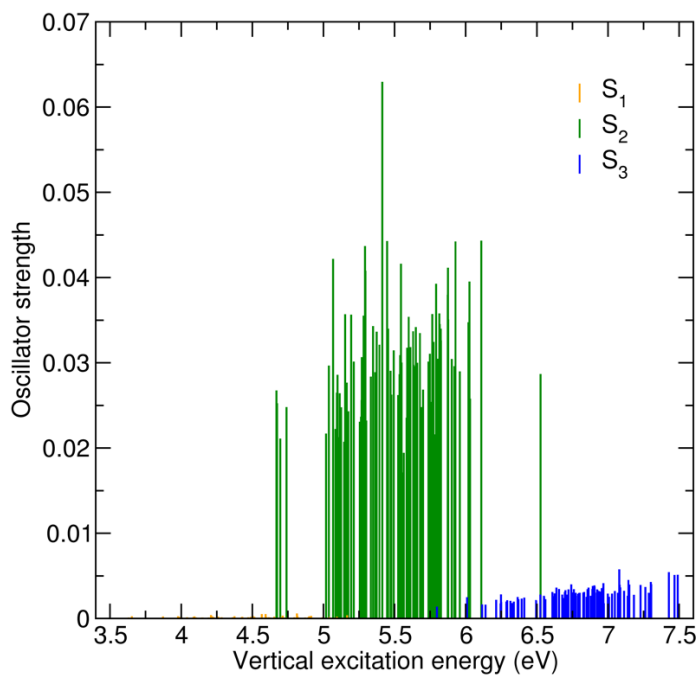

**Figure S10: Oscillator strength of vertical transitions from the $S_0$ to the $S_1$, $S_2$, and $S_3$ states.** Each stick represents a vertical excitation and corresponding oscillator strength for one of the 100 Wigner sampled molecular geometries, computed from the ground state to a given excited state ($S_1$, $S_2$, or $S_3$). Note that the contribution from $S_1$ between 4.5 and 5 eV is so small that it is barely visible in the figure.

*S2.5 Validation of the initial conditions for the ab initio molecular dynamics simulations*

The AIMD simulations were initiated from the nuclear coordinates and with the momenta given by the TSH trajectories once they had reached the $S_0$ state and departed from the region of the $S_1/S_0$ MECI. Figure S11 shows two representative TSH trajectories (grey dots) reaching the ground state. The energy gap between the $S_0$ and $S_1$ states increases shortly after the TSH trajectory hopped to $S_0$, offering a validation for the approximation that the subsequent dynamics on $S_0$ will be considered as adiabatic, *i.e.*, within the Born-Oppenheimer approximation. The percentages in Fig. S11 indicate the contribution of the closed-shell ground-state configuration to the $S_0$ electronic wavefunction for different molecular geometries selected along each trajectory (each such geometry is indicated by a cross). This percentage increases rapidly when the TSH trajectory is driven by the $S_0$ PES. The arrows in Fig. S11 indicate the points selected to initialize the AIMD simulations – the nuclear coordinates and velocities obtained from the TSH trajectory were used to initiate the AIMD.

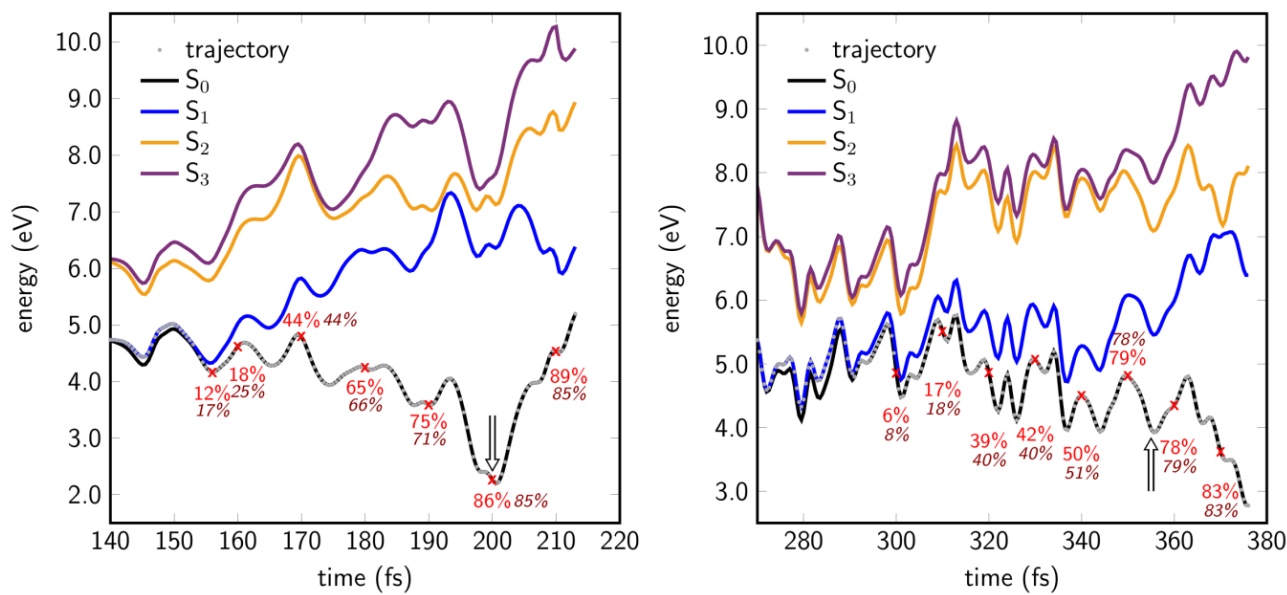

**Figure S11: Electronic energies along two different TSH trajectories.** The dotted line indicates the electronic state driving the TSH dynamics at a given time. The percentages give the contribution of the closed-shell ground-state configuration to the $S_0$ electronic wavefunction (values determined at the SA(4)-CASSCF(10/8) and XMS(4)-CASPT2(10/8) levels of theory are shown in red and in burgundy (italic), respectively) at a given geometry (highlighted by a cross). The arrows show the points used to initiate the subsequent AIMD on $S_0$.

## S2.6 AIMD trajectories out to t = 2 ps and a trajectory exhibiting interconversion between different photoproducts

One focus of the present work was to determine the distribution of the $IP_{vert}$ values for the $S_0 \rightarrow D_0$ transitions of thiophenone and its photoproducts; possible consequences of the energetic proximity of $D_1$ are considered in section S2.9. Each AIMD trajectory starts from a slightly different time (defined relative to the initial photoexcitation process), since the TSH trajectories reach the $S_0$ state at different times. Defining a time-average of the $IP_{vert}$ values over all trajectories is thus challenging, but this does not preclude investigation of the narrowness of the BE distribution observed at later $\Delta t$ (*i.e.* following non-radiative decay to the $S_0$ state) as shown in Fig. 4 of the main text.

Figure S12 shows the $IP_{vert}$ values calculated along an AIMD trajectory in the $S_0$ state, chosen as it shows interconversion between three photoproducts: 2-(2-thiiranyl)ketene (P3), thiophenone, and 2-(2-sulfanylethyl)ketene (P2); the molecular structures are shown below. The $IP_{vert}$ value is computed every 10 fs along the trajectory, and the orange line shows a running average of $IP_{vert}$ over 10 time steps. As with the histograms presented in Fig. 4 of the main text, this plot serves to highlight the similar $IP_{vert}$ values of the different photoproduct configurations (apart from thiophenone itself, which exhibits a slightly higher $IP_{vert}$ value). In all AIMD simulations, compounds were identified based on characteristic atomic connectivities determined by measuring bond lengths or angles, as depicted in Fig. S13.

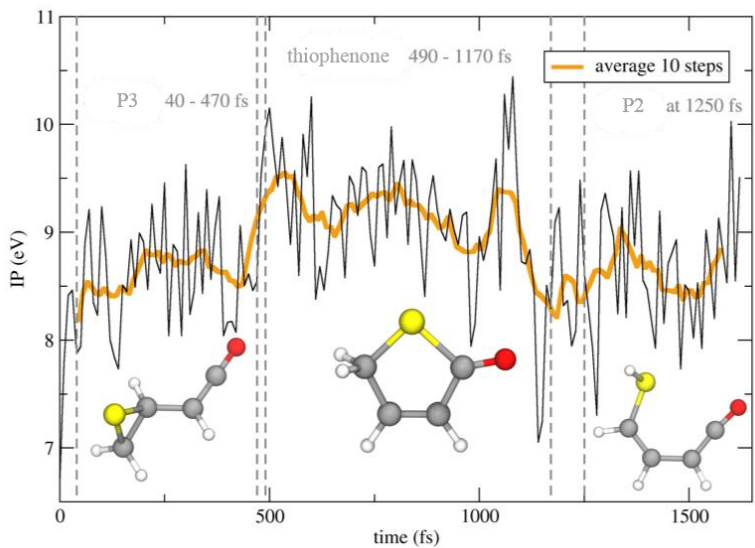

**Figure S12: Vertical ionization potential value along an example AIMD trajectory on $S_0$.**

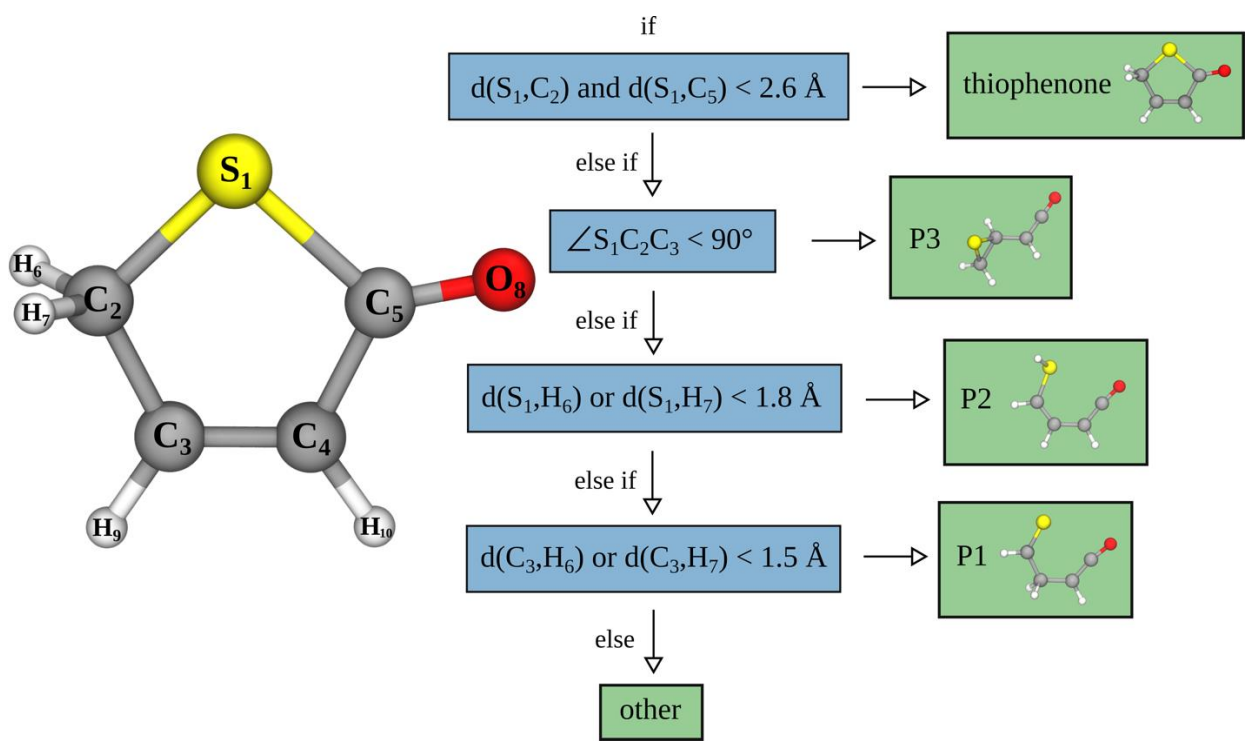

**Figure S13: Decision tree for the identification of photoproducts during the AIMD.**

*S2.7 Normal distribution of IPs*

The histogram of the IP data computed along all AIMD trajectories (without distinguishing the photoproduct) was fitted to a normal distribution, as shown in Fig. S14. The resulting mean value was 8.93 eV with a standard deviation σ = 0.41 eV. The FWHM was calculated from the standard deviation according to $FWHM = 2\sqrt{2\ln(2)}\,\sigma$.

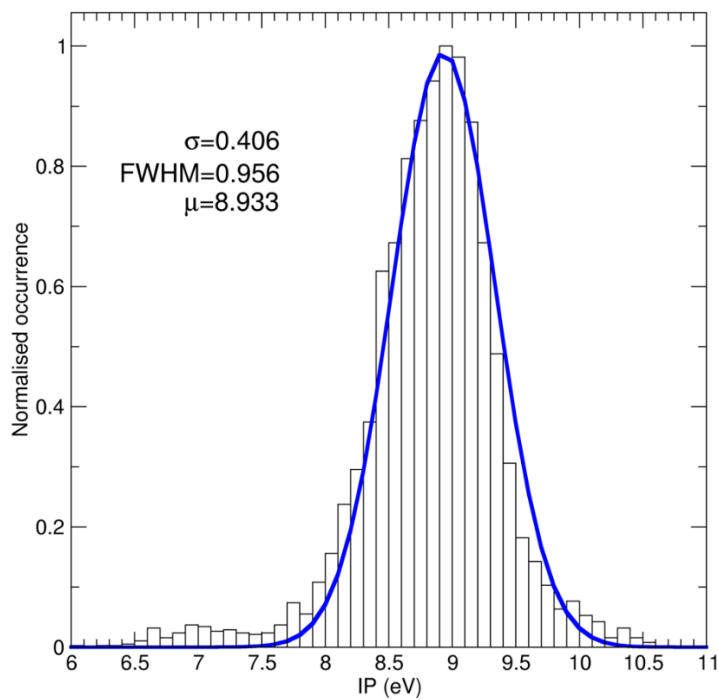

**Figure S14: Histogram of IPs computed along the AIMD trajectories and their fit to a normal distribution.**

*S2.8 Validation of the methodology to compute ionization potentials between $S_0$ and $D_0$*

Table S3 illustrates the close agreement between the IP values computed with CCSD(T)-F12/cc-pVDZ-F12 and MP2-F12/cc-pVDZ-F12 for different possible photoproducts of thiophenone. All geometries for these benchmarking calculations were optimized at the MP2/6-311+G** level of theory. The table also shows energy differences between the electronic ground state energy of the photoproduct and the parent thiophenone molecule. Convergence of the CCSD(T)-F12 and MP2-F12 results with respect to basis set size has been tested by comparing the results obtained with cc-pVDZ-F12 and cc-pVTZ-F12 for thiophenone and for the P3 isomer. Only minor differences were observed between the two basis sets: for thiophenone, the variation in IPs between the two basis sets was 0.05 eV (CCSD(T)-F12) and 0.02 eV (MP2-F12). The level of agreement observed between CCSD(T)-F12 and MP2-F12 for IPs and the convergence observed for both methods with the cc-pVDZ-F12 basis validate the use of MP2-F12/cc-pVDZ-F12 for the results presented in the main text.

Note that the present calculations use a *vertical* approximation to derive the $S_0$–$D_0$ energy gap, *i.e.,* we do not use the lowest energy point on the cation PES, but the point of the cationic PES corresponding to the particular nuclear arrangement in the (neutral) ground state. Both the number of configurations sampled (>4000) and the conformational flexibility of the photoproducts would challenge an estimation of the vibrational overlap based on the harmonic approximation (as is commonly employed).

It is also worth commenting on the possible role of other cationic states during the AIMD process, particularly $D_1$. The analysis presented here was limited to the distribution of the lowest IP band (*i.e.* $S_0^\#$ to $D_0$) during the athermal dynamics following the nonadiabatic relaxation to the ground state. As expected from the LIICs presented above, the $D_1$ cationic state is expected to be only slightly higher in energy than $D_0$. To gain some estimate of the influence of $D_1$ on the high energy side of the IP distribution (*i.e.* of any broadening due to $D_1$), the $D_1$–$D_0$ energy splitting was computed (using XMS(2)-CASPT2(9/8)) for the three photoproducts observed during the AIMD, at their optimized ground state energy. The $D_1$–$D_0$ energy gaps derived in this way are only 0.324 eV (for P1), 1.537 eV (for P2) and 0.493 eV (for P3), suggesting that proper inclusion of the $D_1$ state would likely affect the high energy side of the computed IP distributions.

**Table S3: Ionization potential of thiophenone and photoproducts.** Comparison between CCSD(T)-F12 and MP2-F12 calculated values of the IPs of different photoproducts and of their electronic energies relative to that of the $S_0$ state of thiophenone.

|  | thiophenone | P1 | P2 | P3 | P4 | P5 | P6 |
|---|---|---|---|---|---|---|---|
|  | 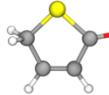 | 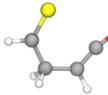 | 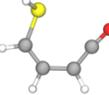 | 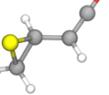 | 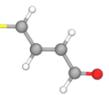 | 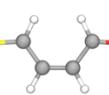 | 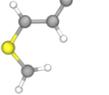 |
| **CCSD(T)-F12** | | | | | | | |
| IP / eV | 9.663 | 9.177 | 8.310 | 8.796 | 9.206 | 9.223 | 7.272 |
| relative to thiophenone | | | | | | | |
| $\Delta S_0$ / eV | | 1.549 | 1.410 | 1.455 | 1.274 | 1.396 | 2.878 |
| $\Delta$ IP / eV | | -0.486 | -1.353 | -0.867 | -0.458 | -0.440 | -2.391 |
| **MP2-F12** | | | | | | | |
| IP / eV | 9.799 | 9.362 | 8.368 | 8.668 | 9.374 | 9.392 | 7.566 |
| relative to thiophenone | | | | | | | |
| $\Delta S_0$ / eV | | 1.567 | 1.417 | 1.367 | 1.386 | 1.512 | 2.761 |
| $\Delta$ IP / eV | | -0.436 | -1.431 | -1.130 | -0.425 | -0.407 | -2.233 |

Finally, the chosen computational strategy for estimating photoelectron spectra was benchmarked by comparing the measured (He I) photoelectron spectrum of thiophenone with that computed using configurations extracted from a ground-state AIMD initiated from a set of nuclear coordinates and momenta sampled from a Wigner distribution (spectrum labelled 'cold' in Fig. S15). Fig. S15 also compares the photoelectron spectrum computed for this 'cold' thiophenone sample with that obtained for $S_0^\#$ thiophenone molecules (*i.e.* molecules that have undergone photoexcitation and subsequent non-adiabatic coupling to the $S_0$ PES, labeled 'hot').

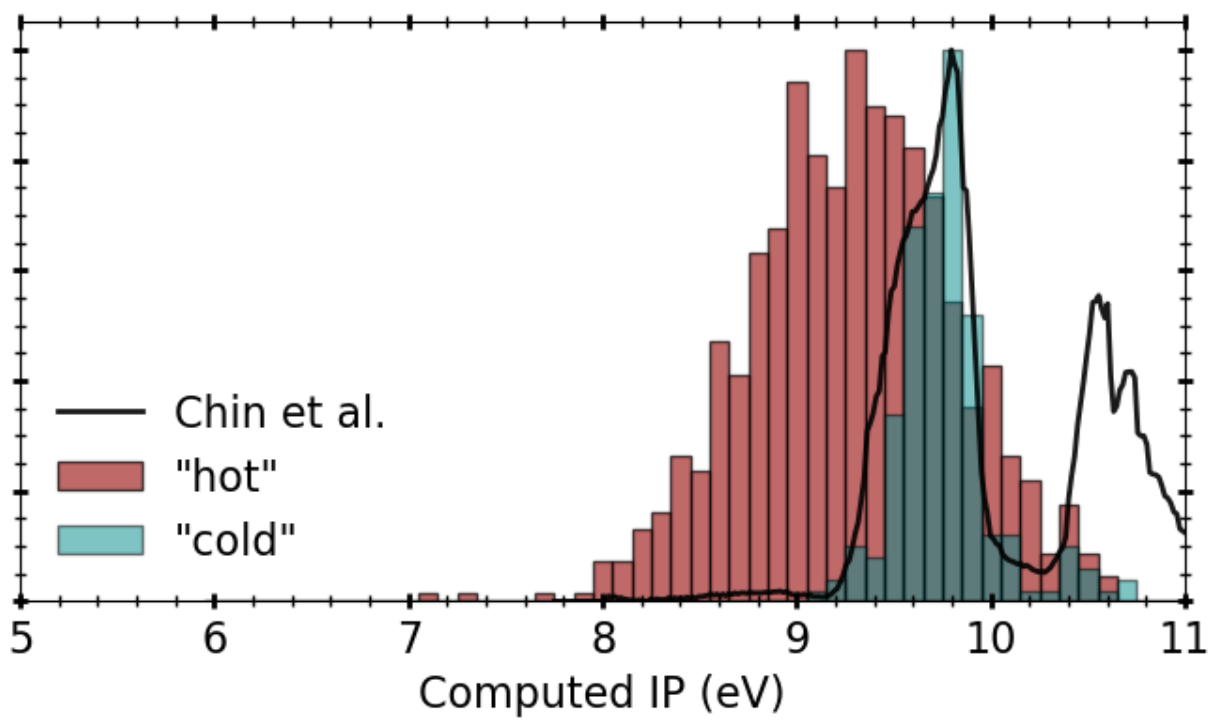

**Figure S15: Comparison of measured (He I) photoelectron spectrum of thiophenone (Chin *et al.*, 1998, solid black trace) and the distributions of computed $S_0 \rightarrow D_0$ IP$_{vert}$ values for 'cold' (turquoise) and 'hot' (burgundy) thiophenone molecules.** The former histogram, obtained from AIMD simulations of $S_0$ molecules with internal energy equal to the zero-point energy only, has been scaled vertically to match the experimental spectrum and illustrates the good agreement with the experimental IP. The latter histogram, which shows the distribution of IP$_{vert}$ values associated with closed-ring 'hot' thiophenone species (computed at the MP2/cc-pVDZ-F12 level of theory) produced after photoexcitation and subsequent relaxation, has been scaled to have the same maximum. Note that the present theory has not attempted to model the $S_0 \rightarrow D_2$ ionization responsible for the BE >10.4 eV peak in the experimental spectrum.

### *S2.9 Nature of the ionization process*

For the electrons identified as being involved in the ionization process, spin densities were computed (at the MP2/cc-pVDZ level of theory) for the optimized structures of the molecules observed during the AIMD; thiophenone, and the ketene photoproducts P1, P2 and P3 observed in the AIMD calculations.

Figure S16 shows the spin densities plotted with an isosurface of 0.02. The spin densities are shown in blue (the green isosurface shows the negative contributions and offers a measure of spin contamination). The spin densities for thiophenone, P1 and P3 are mainly located on the sulfur atom, consistent with the previous conclusion that ionization occurs mainly from the *n*(S) orbital. In the case of P2 (bottom right structure in Fig. S16), the spin density is mostly on the C=C double bond, suggesting that ionization occurs from this π orbital. However, we note that the optimized geometry of the P2 structure may not be fully representative of the hot ground-state photoproduct. For the other photoproducts, the histograms of IPs calculated for the structures observed during the AIMD are all centered at the IP value calculated for the optimized point. For P2, the calculated IP at the optimized point is 8.37 eV, whereas the histogram of IPs of the hot photoproduct is centered at ~8.7 eV. Additionally, the distribution of IPs for P2 is broader than that for any of the other structures. To this end, spin densities were computed for the frames of an AIMD trajectory showing formation of P2. We observe that, for such $S_0^{\#}$ geometries, the spin density is localized not only around the C=C bond but also on the sulfur atom. Thus we conclude that the shift to a higher IP and the broadening of the IP distribution during the AIMD (*cf.* the IP obtained from the ground-state optimized structure) can be understood if we assume that ionization to form $D_0$ occurs from the $\pi$ and the $n(S)$ orbitals. Finally, we note that the characters of the $D_0$ photoproducts accord with that observed in the previously discussed XMS(2)-CASPT2(9/8) calculations.

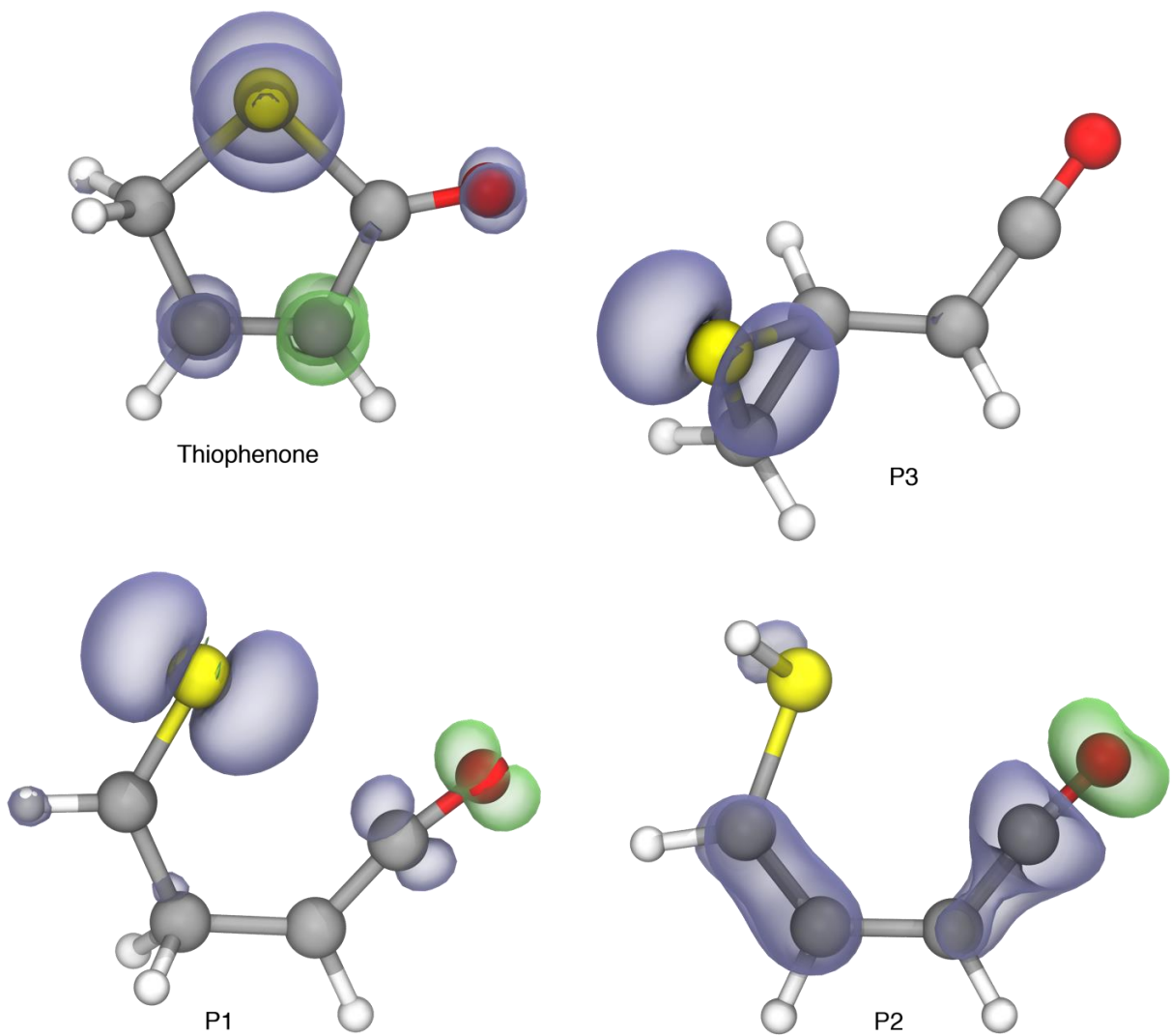

**Figure S16: Spin densities of thiophenone and photoproducts.** Spin densities (blue) of the optimized structures of the photoproducts observed, and their negative contributions (green).

### *S2.10 AIMD trajectories out to t = 100 ps and trajectories exhibiting dissociation to CO + thioacrolein*

To explore the possible evolution of the ground state photoproducts over longer timescales, 10 AIMD trajectories (out of the initial 22) were propagated until $t = 100$ ps. Issues of particular interest here were (i) whether the 'hot' $S_0^\#$ thiophenone photoproducts would eventually ring-open, and (ii) whether the primary photoproducts might undergo further fragmentation. At the start of the long-time AIMD simulations (*i.e.*, at $t = 2$ ps), the chosen pool of 10 molecules involved 4 (*i.e.* 40%) with 'hot' thiophenone structures, 4 P1 and 2 P2 structures. We stress that these proportions were chosen solely for the purpose of investigating the possible fate of some specific photoproducts (particularly 'hot' thiophenone) during the long-time exploratory AIMD; they do not reflect the relative photoproduct yields after $t = 2$ ps (recall Fig. 4 in the main paper). Figure S17 shows how these photoproduct proportions evolve over the full length of the AIMD. These (admittedly small number of) trajectories show the population of thiophenone decreasing and new dissociation products (thioacrolein + CO) arising from decay of both 'hot' thiophenone and photoproduct P1. The computed $S_0$–$D_0$ IP$_{vert}$ value for thioacrolein (9.04 and 8.99 eV for the Z- and E-isomers, respectively, at the MP2-F12/cc-pVDZ-F12 level of theory) is in very good agreement with the experimental value [*Bock1982*]. Clearly, the present long time AIMD analysis is statistically limited, but it serves to highlight possible pathways that can alter the composition of the photoproducts formed on the $S_0$ PES over time in a manner that could account for the long time variation in the TRPES signal observed in the range $8 \leq \text{BE} \leq 9.5$ eV (Fig. 4(a)).

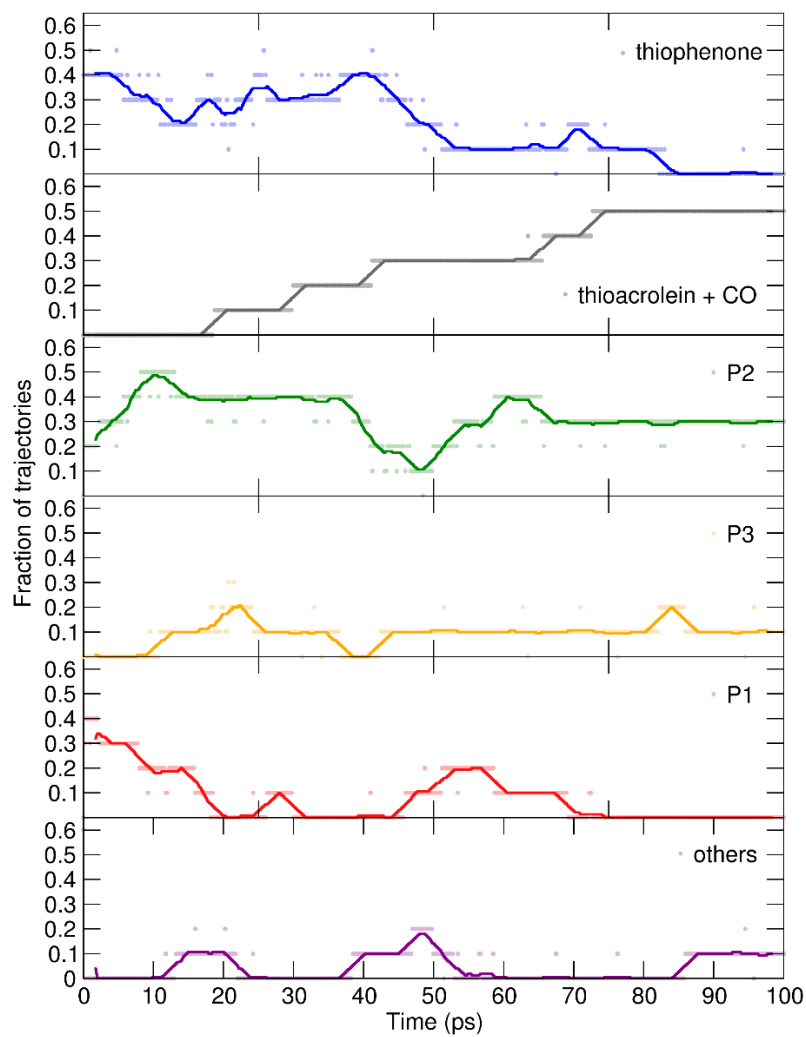

**Figure S17: Photoproduct distribution during the long-time AIMD simulations.** The dots indicate the proportion of a given photoproduct sampled every 0.25 ps, while the solid lines show the corresponding running average (over 15 time steps).

**Supplementary References:**

Arbelo-González, W., Crespo-Otero, R., & Barbatti, M. Steady and time-resolved photoelectron spectra based on nuclear ensembles. *Journal of Chemical Theory and Computation* **12**, 5037-5049 (2016).

Bock, H., Mohmand, S., Hirabayashi, T. & Semkow, A. Gas-phase reactions. 29. Thioacrolein. *J. Am. Chem. Soc.* **104**, 312-313 (1982).

Chin, W.S. et al. He I and He II photoelectron spectra of thiophenones. *J. Electron Spectroscopy Related Phenomena* **88–91**, 97–101 (1998).

Granucci, G. & Persico, M. Critical appraisal of the fewest switches algorithm for surface hopping. *J. Chem. Phys.* **126**, 134114 (2007).

Hattig, C., Klopper, W., Köhn, A. & Tew, D.P. Explicitly Correlated Electrons in Molecules. *Chem. Rev*. **112**, 4−74 (2012).

Kong, L., Bischoff, F.A. & Valeev, E.F. Explicitly Correlated R12/F12 Methods for Electronic Structure. *Chem. Rev*. **112**, 75− 107 (2012).

Mai, S., Marquetand, P. & González, L. Nonadiabatic dynamics: the SHARC approach. *WIREs Comput. Mol. Sci*. **8**, 6-e1370 (2018).

Mai, S. et al. SHARC2.0: Surface Hopping Including Arbitrary Couplings — Program Package for Non-Adiabatic Dynamics, *sharc-md.org*, 2018.

Plasser, F. TheoDORE 2.0: a package for theoretical density, orbital relaxation, and exciton analysis. *Available from http://theodore-qc.sourceforge.netF*.

Plasser, F., Wormit, M., Dreuw, A. New tools for the systematic analysis and visualization of electronic excitations. I. Formalism. *J. Chem. Phys*. **141**, 024106 (2014).

Svetina, C. et al. The Low Density Matter (LDM) beamline at FERMI: optical layout and first commissioning. *J. Synchrotron Rad*. **22**, 538–543 (2015).

Tully, J.C. Molecular dynamics with electronic transitions. *J. Chem. Phys*. **93**, 1061–1071 (1990).